\documentclass[a4paper,fleqn,usenatbib,useAMS]{mnras}
\usepackage{color}
\usepackage{graphicx}	
\usepackage{amsmath}	
\usepackage{amssymb}	
\usepackage{multicol}        
\usepackage{bm}		
\usepackage{pdflscape}	

\def\ltsima{$\; \buildrel < \over \sim \;$}
\def\simlt{\lower.5ex\hbox{\ltsima}}
\def\gtsima{$\; \buildrel > \over \sim \;$}
\def\simgt{\lower.5ex\hbox{\gtsima}}

\title[The origin of assembly bias]{ZOMG - I. How the cosmic web inhibits halo growth and generates assembly bias}
\author[Borzyszkowski, et al.]{Mikolaj Borzyszkowski\thanks{E-mail: mikolajb@uni-bonn.de}, Cristiano Porciani, Emilio Romano-D{\'{\i}}az and Enrico Garaldi\thanks{Member of the International Max Planck Research School (IMPRS) for Astronomy and Astrophysics at the Universities of Bonn and Cologne}\\
Argelander-Institut f\"ur Astronomie, University of Bonn, Auf dem H\"ugel 71, D-53121 Bonn, Germany}
\begin{document}

%
%
%
%


\def\refads@jnl#1{{\rm#1}}

\def\aj{\refads@jnl{AJ}}                   
\def\actaa{\refads@jnl{Acta Astron.}}      
\def\araa{\refads@jnl{ARA\&A}}             
\def\apj{\refads@jnl{ApJ}}                 
\def\apjl{\refads@jnl{ApJ}}                
\def\apjs{\refads@jnl{ApJS}}               
\def\ao{\refads@jnl{Appl.~Opt.}}           
\def\apss{\refads@jnl{Ap\&SS}}             
\def\aap{\refads@jnl{A\&A}}                
\def\aapr{\refads@jnl{A\&A~Rev.}}          
\def\aaps{\refads@jnl{A\&AS}}              
\def\azh{\refads@jnl{AZh}}                 
\def\baas{\refads@jnl{BAAS}}               
\def\bac{\refads@jnl{Bull. astr. Inst. Czechosl.}}
\def\caa{\refads@jnl{Chinese Astron. Astrophys.}}
\def\cjaa{\refads@jnl{Chinese J. Astron. Astrophys.}}
\def\icarus{\refads@jnl{Icarus}}           
\def\jcap{\refads@jnl{J. Cosmology Astropart. Phys.}}
\def\jrasc{\refads@jnl{JRASC}}             
\def\memras{\refads@jnl{MmRAS}}            
\def\mnras{\refads@jnl{MNRAS}}             
\def\na{\refads@jnl{New A}}                
\def\nar{\refads@jnl{New A Rev.}}          
\def\pra{\refads@jnl{Phys.~Rev.~A}}        
\def\prb{\refads@jnl{Phys.~Rev.~B}}        
\def\prc{\refads@jnl{Phys.~Rev.~C}}        
\def\prd{\refads@jnl{Phys.~Rev.~D}}        
\def\pre{\refads@jnl{Phys.~Rev.~E}}        
\def\prl{\refads@jnl{Phys.~Rev.~Lett.}}    
\def\pasa{\refads@jnl{PASA}}               
\def\pasp{\refads@jnl{PASP}}               
\def\pasj{\refads@jnl{PASJ}}               
\def\rmxaa{\refads@jnl{Rev. Mexicana Astron. Astrofis.}}%
\def\qjras{\refads@jnl{QJRAS}}             
\def\skytel{\refads@jnl{S\&T}}             
\def\solphys{\refads@jnl{Sol.~Phys.}}      
\def\sovast{\refads@jnl{Soviet~Ast.}}      
\def\ssr{\refads@jnl{Space~Sci.~Rev.}}     
\def\zap{\refads@jnl{ZAp}}                 
\def\nat{\refads@jnl{Nature}}              
\def\iaucirc{\refads@jnl{IAU~Circ.}}       
\def\aplett{\refads@jnl{Astrophys.~Lett.}} 
\def\apspr{\refads@jnl{Astrophys.~Space~Phys.~Res.}}
\def\bain{\refads@jnl{Bull.~Astron.~Inst.~Netherlands}} 
\def\fcp{\refads@jnl{Fund.~Cosmic~Phys.}}  
\def\gca{\refads@jnl{Geochim.~Cosmochim.~Acta}}   
\def\grl{\refads@jnl{Geophys.~Res.~Lett.}} 
\def\jcp{\refads@jnl{J.~Chem.~Phys.}}      
\def\jgr{\refads@jnl{J.~Geophys.~Res.}}    
\def\jqsrt{\refads@jnl{J.~Quant.~Spec.~Radiat.~Transf.}}
\def\memsai{\refads@jnl{Mem.~Soc.~Astron.~Italiana}}
\def\nphysa{\refads@jnl{Nucl.~Phys.~A}}   
\def\physrep{\refads@jnl{Phys.~Rep.}}   
\def\physscr{\refads@jnl{Phys.~Scr}}   
\def\planss{\refads@jnl{Planet.~Space~Sci.}}   
\def\procspie{\refads@jnl{Proc.~SPIE}}   

\let\astap=\aap
\let\apjlett=\apjl
\let\apjsupp=\apjs
\let\applopt=\ao

\date{\today}

\pagerange{\pageref{firstpage}--\pageref{lastpage}} \pubyear{2016}

\maketitle

\label{firstpage} 

\begin{abstract}
The clustering of dark matter haloes with fixed mass depends on their formation history, an effect known as assembly bias.
We use zoom $N$-body simulations to investigate the origin of this phenomenon. 
For each halo at redshift $z=0$, we determine the time in which the physical volume containing its final mass becomes stable. 
We consider five examples for which this happens at $z\sim1.5$ and two that do not stabilize by $z=0$. 
The zoom simulations show that early-collapsing haloes do not grow in mass at $z=0$ while late-forming ones show a net inflow. 
The reason is that `accreting' haloes are located at the nodes of a network of thin filaments feeding them. 
Conversely, each `stalled' halo lies within a prominent filament that is thicker than the halo size. 
Infalling material from the surroundings becomes part of the filament while matter within it recedes from the halo. 
We conclude that assembly bias originates from quenching halo growth due to tidal forces following the formation of non-linear structures in the cosmic web, as previously conjectured in the literature. 
Also the internal dynamics of the haloes change: the velocity anisotropy profile is biased towards radial (tangential) orbits in accreting (stalled) haloes. 
Our findings reveal the cause of the yet unexplained dependence of halo clustering on the anisotropy.
Finally, we extend the excursion-set theory to account for these effects. 
A simple criterion based on the ellipticity of the linear tidal field combined with the spherical collapse model provides excellent predictions for both classes of haloes.
\end{abstract}


\definecolor{ColorAbu}{rgb}{0.8, 0., 0.8}
\definecolor{ColorAmun}{rgb}{0.5, 0.1, 0.5}
\definecolor{ColorSeth}{rgb}{0.75, 0.75, 0}
\definecolor{ColorSupay}{rgb}{1.0, 0.4, 0.0}
\definecolor{ColorSedna}{rgb}{0.0, 0.75, 0.75}
\definecolor{ColorSobek}{rgb}{0.0, 0.5, 0.0}
\definecolor{ColorSiris}{rgb}{0.,0.,0.}

\defcitealias{Borzyszkowski2014}{BLP} 
\defcitealias{Ludlow2014}{LBP} 

\section{Introduction}
In the standard model of cosmology, galaxies are surrounded by extended haloes made of cold dark matter (CDM).  
These CDM clumps are generated by primordial density perturbations that are gravitationally unstable.
The power spectral density of the fluctuations is such that haloes form hierarchically in a bottom-up fashion.
The first generation of haloes comes into being at redshift $z\sim 60$ with Earth masses \citep*{Diemandetal2005}.
Larger and larger objects are then progressively assembled through halo mergers and accretion of loose material.
$N$-body simulations are the ideal tool to study this highly non-linear process.
They show that the host haloes of present-day bright galaxies start forming at a redshift of a few and become a prominent population around $z\sim 1$. 

Galaxies and their host haloes are embedded in the large-scale structure of the Universe.
Matter is organised in a vast network of knots, filaments and sheets known as the `cosmic web' \citep{BKP}.  
Distinct halo populations trace the various elements of the web in different ways and thus show dissimilar clustering properties \citep{Hahnetal2007a, Hahnetal2007b, AragonCalvoetal2007, Cautunetal13}.
For a long time, numerical studies have focussed on determining the clustering amplitude (e.g. the amplitude of the two-point correlation function) as a function of halo mass (see also Section~\ref{sec_model} for a more theoretical perspective).
More recently, however, it has been shown that, at fixed mass, halo clustering depends on the formation time (\citealt*{Gao2005}; \citealt{Wechsleretal06, Zhuetal06}; \citealt*{LiMoGao2008}; \citealt{Wangetal2011, Sunayamaetal16}) and other correlated variables (\citealt{Wechsleretal06, Jingetal07, Gao2007, Bett2007}; \citealt*{Anguloetal08}; \citealt{Faltenbacher2010, Lieetal2013}). 
For galaxy-sized objects (the subject of this paper), haloes that formed earlier cluster more strongly than their `younger' counterparts. 
In other words, older haloes tend to reside in denser environments \citep{ShethTormen04, Harkeretal06}. 
This set of phenomena is known under the collective name of `halo assembly bias'. 
It is an intriguing open question whether also `galaxy assembly bias' exists, i.e. to what extent the assembly history of the host haloes influences galaxy properties and imprints detectable signatures in the galaxy-clustering pattern (e.g. \citealt*{Jungetal14,Hearinetal15}; \citealt{Hearinetal16, Linetal16}).

This article is the first in a series of three describing a numerical project named Zooming On a Mob of Galaxies (ZOMG).
The aim of the ZOMG simulations is to study how cosmic environment regulates the growth and the structure of dark matter haloes and of their central and satellite galaxies, hence to investigate the physical origin of halo and (possibly) galaxy assembly bias. 
Here we focus on the formation of dark matter haloes and therefore use high-resolution $N$-body simulations. 
In the companion papers, instead, we also consider the baryonic component and use hydrodynamic simulations including star formation and feedback.
In particular, in paper II  \citep{PaperII}, we investigate how environment regulates the gas supply and the properties of the central galaxies while, in paper III \citep{PaperIII}, we shift our attention to the satellite galaxies. 

In this work, we study the formation and evolution of the host haloes of present-day $L_*$ galaxies.
These haloes are of particular interest because they are the sites where star formation is most efficient and the stellar-to-halo mass ratio peaks \citep*{Mosteretal, Behroozi}. 
In $N$-body simulations, they also present a number of puzzling properties which are interconnected. 
First, they show a strong assembly bias.  
Second, the bulk of them seem to have assembled early on \citep*[][hereafter \citetalias{Borzyszkowski2014}]{Borzyszkowski2014} and, consistently, do not present signs of mass growth at later times (\citealt*{Diemandetal2007}; \citetalias{Borzyszkowski2014}). 
Third, the region within which they are dynamically stable (i.e. where the mean radial velocity of matter vanishes) appears to be significantly more extended than expected from simple theoretical considerations \citep{Pradaetal2006}. 
Fourth, given the linear overdensity and the tides in the Lagrangian patch from which they originate, they form much earlier than predicted by popular collapse models (\citealt*[][hereafter \citetalias{Ludlow2014}]{Ludlow2014}; \citetalias{Borzyszkowski2014}; \citealt{HahnParanjape14}).   
The main motivation for our work is to verify these results and provide a comprehensive and coherent explanation for all of them.

Several studies indicate that halo assembly bias can be explained in terms of a population of haloes that stop accreting matter once they find themselves in dense environments (\citealt*{Wang2007}; \citealt{Dalal2008,Hahnetal2009,Wangetal2011}). 
However, the detailed physical origin of this phenomenon has not been clarified yet. 
Although older haloes lie closer to more massive objects \citep{Wang2007}, mass loss due to tidal stripping during fly-bys does not seem to play any relevant role in establishing the bias \citep{Wang2007, Hahnetal2009, Wangetal2011}.  
\citet{Wang2007} note that the dark matter flow surrounding the older haloes is hotter and might escape capture from the halo potential well. 
\citet{Hahnetal2009} provide evidence in favour of tidally suppressed accretion driven by the strong velocity shear within marked filaments of the large-scale structure. 
Consistently, \citet{Wangetal2011} find that halo formation time at fixed mass correlates with the strength of the local tidal field. 
All these valuable investigations are based on fairly large cosmological $N$-body simulations providing substantial statistical samples but with relatively poor spatial and mass resolution.  
For this reason they are ideal for spotting correlations in the data although they give little insight into the physics of assembly bias.
In fact, there are questions that cannot be addressed with simulations in which galaxy-sized haloes are resolved with $10^2-10^4$ particles leading to a very poor sampling of the phase-space configuration of the halo and its surroundings.
Information on the internal structure of the haloes as well as the precise geometry with which matter accretes on to them can only be inferred stacking thousands of objects with similar apparent characteristics. 
In this way, one always ends up mixing information coming from different objects in more or less disparate dynamical states.  The precise outcome of
this procedure relies on assumptions (for instance a stacking criterion) and can provide only generic indications or trends. 
Actually, the corresponding results cannot be treated as facts at the level of individual haloes until they are corroborated with further, direct evidence. 
Complementary to this line of research, we use the zoom  technique to study a small sample of representative haloes at very high resolution.
This enables us to investigate the accretion-pattern of individual haloes in great detail and identify the series of events leading to halted growth.
Furthermore, our simulations allow us to make a connection between matter accretion from the environment and the dynamical structure of the haloes.
This is particularly interesting since assembly bias has been found to be surprisingly strong when haloes are split based 
on the velocity anisotropy parameter \citep{Faltenbacher2010}.
Our study reveals the unknown origin of this signal and its connection with the halted halo growth.

This paper is structured as follows. 
In Section~\ref{sec_sim} we introduce our numerical simulations and lay out the methods we use to analyse them. 
In Section~\ref{sec_results} we present our results and clarify the origin of halo assembly bias. 
We show that there is a tight link between halo assembly and the cosmic web: the accretion rate of matter on to galaxy-sized haloes is regulated by the environment.
In particular, mass growth is inhibited for haloes embedded in prominent filaments due to the strongly sheared velocity field.
Furthermore, the characteristic accretion pattern determines the internal dynamics of the haloes quantified in terms of the velocity anisotropy profile.
In Section~\ref{sec_model} we discuss the mass-accretion history of our haloes in the context of the excursion-set model \citep{BCEK, Zentner}.
We show that the most common implementations of this theoretical framework fail to capture the effects of the inhibited mass growth in filaments.
Hence we propose a simple extension to the model that nicely accounts for them.  
Finally, we draw our conclusions in Section~\ref{sec_conc}.

\section{Simulations and methods}\label{sec_sim}
We use the tree-PM $N$-body code  {\sc pgadget-3} \citep{Springel2005} to simulate the formation and evolution of dark matter haloes in the $\Lambda$CDM cosmology.
Consistently with the most recent observations \citep{PlanckCosmology2014},
we assume the cosmological parameters $\Omega_\Lambda=0.692$, $\Omega_{\rm m}=0.308$, $\Omega_{\rm b}=0.0481$ and $H_0=100\,h~{\rm km\,s^{-1}Mpc^{-1}}=67.8~{\rm km\,s^{-1}Mpc^{-1}}$. 
Initial conditions are set at $z=99$ 
using second-order Lagrangian perturbation theory with the {\sc music} code \citep{HahnAbel11}.
The linear power spectrum of the density fluctuations 
\citep{Eisenstein-Hu} is characterised by the scalar spectral index $n=0.9608$ and the normalisation amplitude $\sigma_8=0.826$.

We first carry out a simulation within a periodic cube of side 50 $h^{-1}$ Mpc at a uniform initial resolution
in which the matter distribution is sampled with $512^3$ equal mass particles (hereafter U512).
This box forms the parent run from which we select sub-volumes to sample at higher resolution using
the multimass zoom technique (Section~\ref{D-ZOMG}).
To study the stability of our results with respect to numerical discretization, 
we run a second simulation with uniform initial resolution
starting from the same initial conditions but containing only $256^3$ identical particles (U256).

Using the {\sc amiga halo finder} \citep*{Gilletal2004,KnollmannKnebe09},
we conventionally identify dark matter haloes as spherical regions enclosing a mean matter density equal to 200 times the critical density of the universe (the consequences of adopting this common definition will be discussed later). Additionally, we apply an unbinding procedure which iteratively purges the halo of the particles that move faster than $1.5$ times the escape velocity. We define the halo radius $R_{\rm h}$ as the radius of the smallest sphere enclosing the bound particles and the halo mass $M_{\rm h}$ as their total mass.

\begin{figure}
\includegraphics[width=0.49\textwidth,bb=0 0 677 567,keepaspectratio=true]{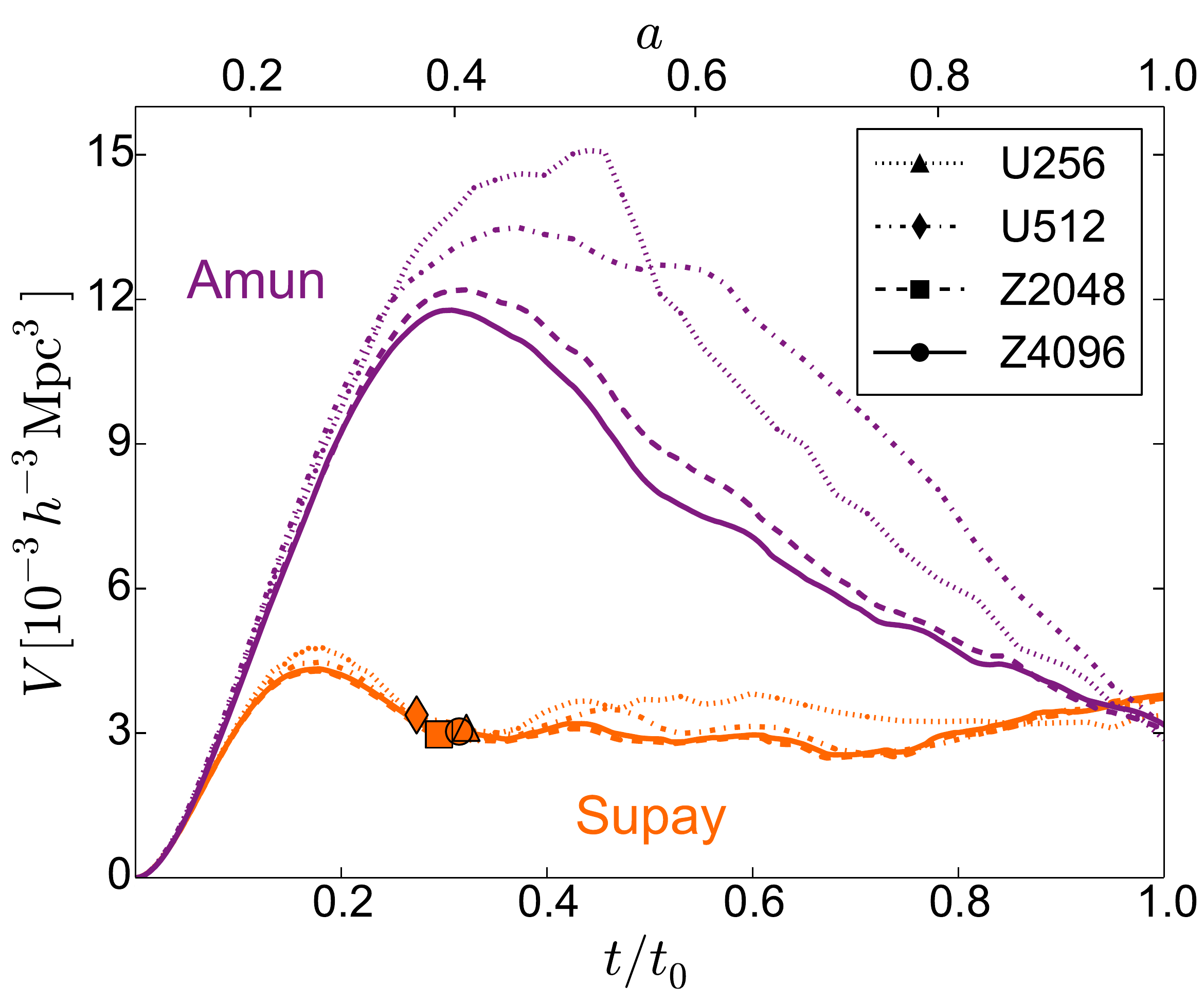}
\caption{The volume of the collapsing patch that ends up forming a dark matter halo is plotted as a function of time. Results are shown for two haloes with approximately the same mass 
$M_{\rm h}\sim 4\times 10^{11} h^{-1} {\rm M}_\odot$.
Different line styles correspond to $N$-body simulations with different mass and force resolution and the
solid symbols indicate the collapse time, $t_{\rm c}$, defined as in equation~(\ref{eq_collapsetime}).
The top series of curves refers to Amun, a halo whose outer shells are still contracting at the present time and
that represents the prototypical `accreting' halo.
On the other hand, the bottom set shows results for Supay, a representative of the class of `stalled' haloes.
Supay turned around earlier than Amun and reached a much lower maximum volume. 
Although Supay has been identified at the present time,
its volume has been stable for $\sim70$ per cent of the life of the universe.
Note that the measurement of $t_{\rm c}$ is very robust with respect to the numerical resolution of the simulations.}
\label{fig_volume_example}
\end{figure}

\subsection{Collapse time}\label{sec_age}
Halo formation can be discussed either in terms of the growing mass of the most massive progenitor
or as the coherent collapse of a single perturbation with the final halo mass.
The first approach is popular in numerical studies while the second one characterises many analytical models. 
In this work, we follow both methods and make use of a particular definition of halo collapse time first introduced by \citetalias{Borzyszkowski2014} in order to establish a direct correspondence between the predictions
of theoretical models and $N$-body simulations. 

To begin with, we compute the (time-dependent) physical volume of the collapsing patch out of which a halo forms. This goal is achieved implementing the following procedure \citepalias[for further details see][]{Borzyszkowski2014}: 
(i) a halo is identified at some epoch (we use the present time, $t_0$); (ii) the halo particles are tagged and their inertia tensor 
is evaluated for every snapshot of the simulation (with respect to their centre of mass); (iii) an ellipsoid with principal axes oriented as the eigenvectors of the inertia tensor and shape determined by the ratios of the square roots of the eigenvalues
is located at the centre of mass of the particle set; (iv) the ellipsoid is rigidly rescaled until it encloses the 
final mass of the halo. The volume of the ellipsoid defines the volume of the collapsing patch, $V(t)$.

In Fig.~\ref{fig_volume_example} we show results obtained for two haloes with approximately the same mass named Amun and Supay (various line styles correspond to simulations with different mass resolutions). 
Qualitatively the time evolution of $V$ conforms to the predictions of the spherical-collapse model. The volume first increases in a decelerated fashion due to the combined action of cosmic expansion and self gravity. At a certain point the perturbation turns around and from this moment onwards $V$ decreases until `virialization' is achieved (meaning that $V$ stabilizes around a constant value).
However, in spite of these similarities, the patches that give origin to Amun and Supay show very different evolutionary paths.
Supay turns around earlier (when its mean physical density is comparable with the present-day one)
and virializes shortly after turnaround.
On the other hand, Amun has a much lower density at turn around\footnote{
For both Amun and Supay, the ratio between the maximum of $V(t)$ and $V(t_0)$ is much lower than the factor of eight
one would expect based on the spherical collapse model and simple descriptions of the virialisation process
\citep[similar findings have been presented by][]{Diemandetal2007}.} and has not yet reached a stable configuration by the present time.
Note, that the final volume is determined by the fixed density threshold with which haloes have been defined.

\begin{figure}
 \centering
 \includegraphics[width=0.49\textwidth,bb=0 0 686 530,keepaspectratio=true]{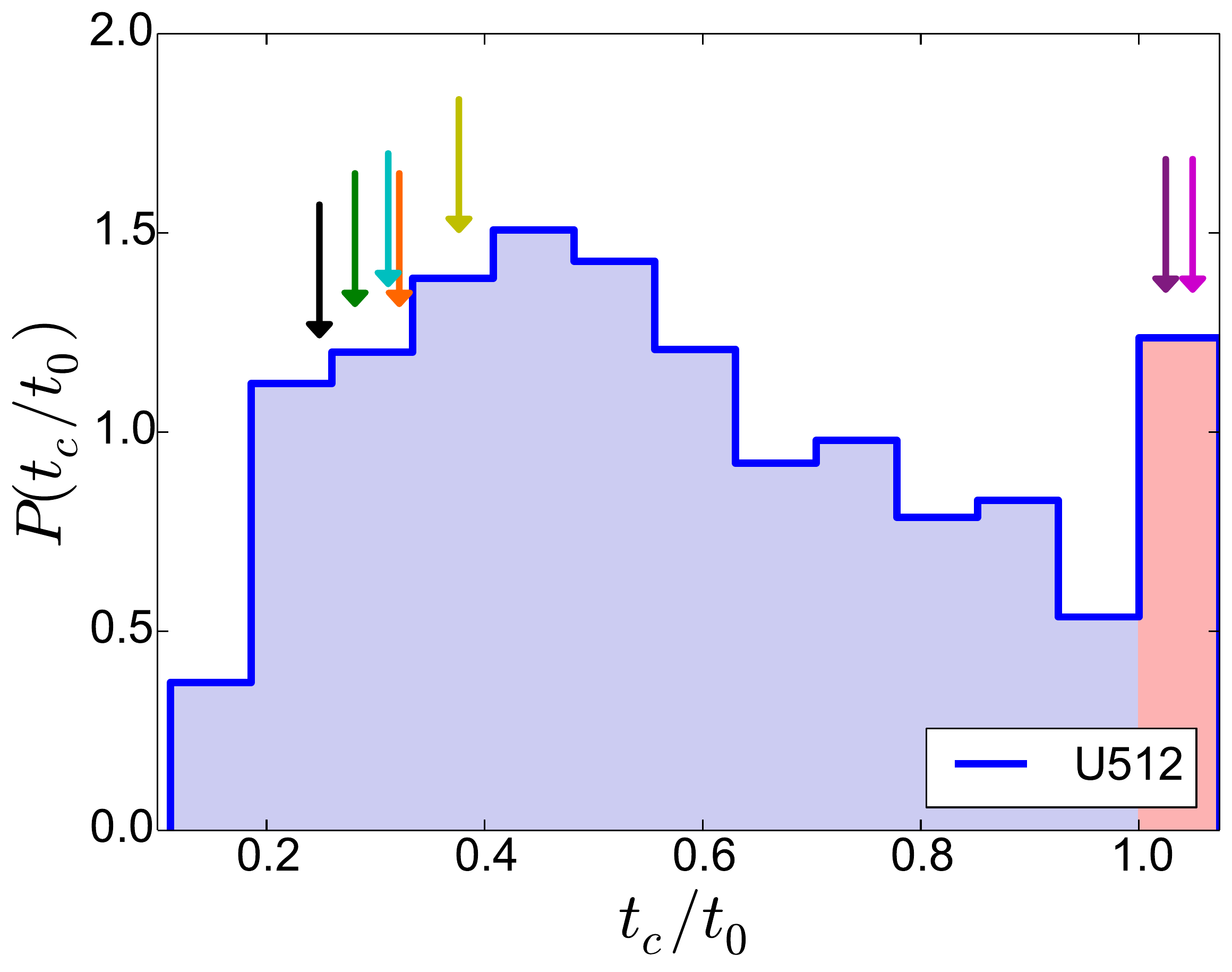}
 \caption{Probability distribution of collapse times for haloes with $2.0<M_{\rm h}/(10^{11}h^{-1}{\rm M}_\odot)<20.0$ extracted from the U512 simulation.
 The spike in the rightmost bin (painted red) represents the total contribution of haloes with $t_{\rm c}\gtrsim t_0$. 
The vertical arrows indicate the collapse time of the haloes selected for re-simulation.}
\label{fig_colltime_dist}
\end{figure}
\begin{figure}
 \centering
 \includegraphics[width=0.49\textwidth,bb=0 0 684 562]{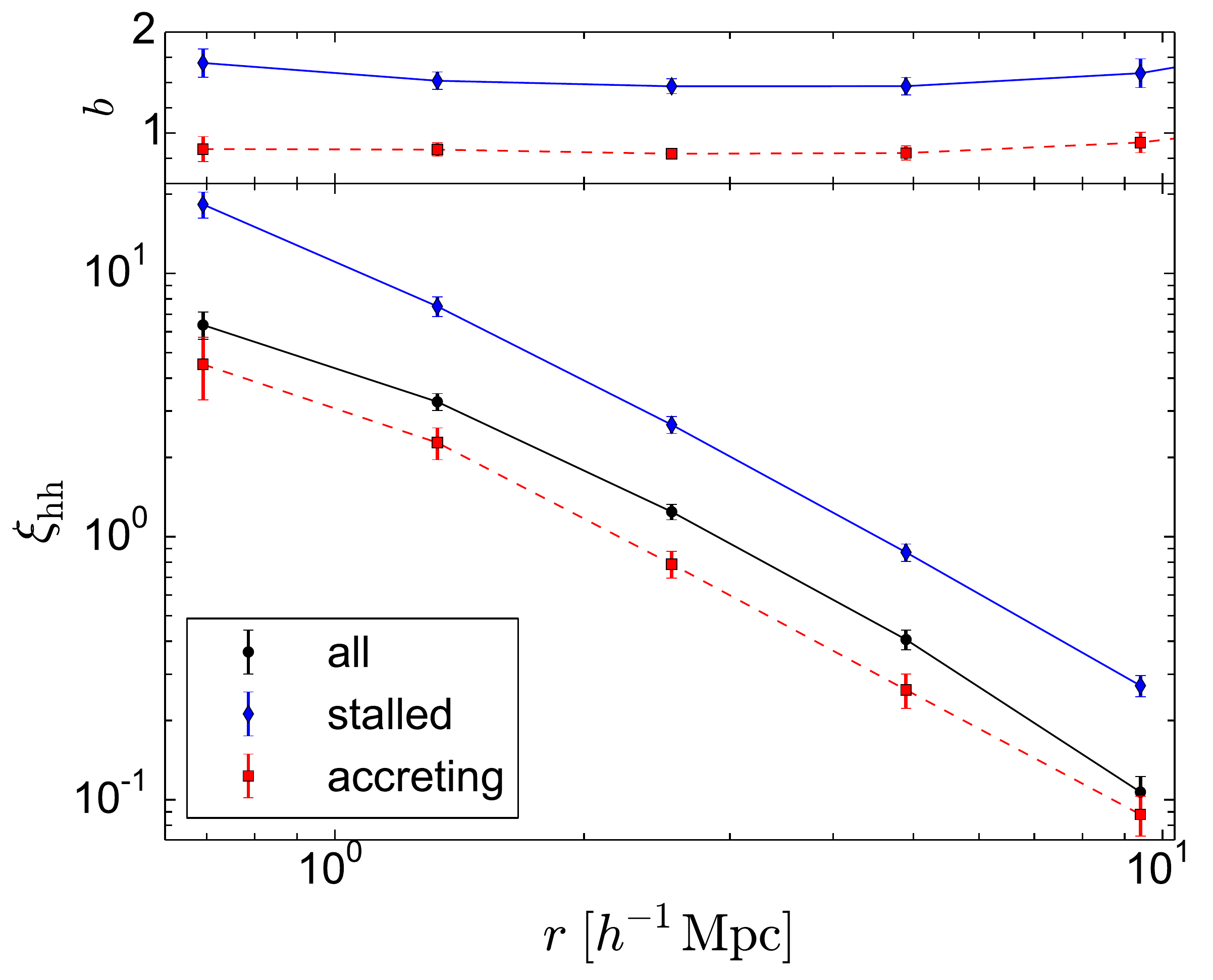}
 \caption{Bottom: data points show the two-point correlation function of haloes with mass
 $2.0<M_{\rm h}/(10^{11} h^{-1}{\rm M}_\odot)<20.0$ and that populate
 the lower (red) and upper (blue) 20 per cent tails of the collapse-time distribution. 
 For comparison, the correlation function of all haloes in the mass bin is shown in black.
 All data come from the U512 simulation. Error bars are obtained bootstrapping the halo positions.
 Top: halo assembly bias defined as the square root of the ratio between the two-point correlations
 of a subpopulation and of the full sample.}
 \label{fig_hhcorr}
\end{figure}

It is tempting to classify haloes based on these marked differences.
We thus define the collapse time of a halo, $t_{\rm c}$, as the moment in which $V(t)$ becomes stable. 
In order to account for the late-time oscillations of $V(t)$, we determine $t_{\rm c}$ as the earliest epoch for which
\begin{eqnarray} \label{eq_collapsetime}
 \int_{t_{\rm c}}^{t_{\rm max}}\left[V(t)-V(t_{\rm c})\right]\,{\rm d}t=0\;,
 \label{eqdeftc}
\end{eqnarray}
with $t_{\rm max}=t_0$ (i.e. the final time of our simulations although, in principle, it would make sense 
to take the limit $t_{\rm max}\to +\infty$).
Solving this equation, for Supay we find that $t_{\rm c}\simeq 0.31\, t_0$ (or, equivalently, $a_{\rm c}\simeq
0.4$ in terms of the cosmic expansion factor normalized to unity at the present time).
For haloes like Amun that have not yet collapsed, instead, we can only provide a lower limit $t_{\rm c}>t_0$
[although a precise value for $t_{\rm c}$ could be determined by running the simulations into the future
this would not influence the results of our paper]. 

Fig.~\ref{fig_colltime_dist} shows the probability distribution of $t_{\rm c}$ for haloes of mass
$2.0<M_{\rm h}/(10^{11} h^{-1}{\rm M}_\odot)<20.0$ in the U512 simulation.
The median value is $0.54\, t_0$ and the scatter around it is quite large. The distribution is skewed towards large values of $t_{\rm c}$. The earliest-forming object collapsed at $t_{\rm c}=0.11\, t_0$ while approximately 9 per cent of the haloes have not collapsed by the present time.

Halo assembly bias can be clearly detected after partitioning objects of a given mass in different classes based
on $t_{\rm c}$. 
The red and blue lines in Fig.~\ref{fig_hhcorr} show the two-point correlation functions of the haloes with the 20 per cent earliest ($t_{\rm c}<0.33 \,t_0$) and latest ($t_{\rm c}>0.85\, t_0$) collapse times, respectively. For comparison, the black line indicates the autocorrelation function of all haloes in the same mass range.
The size of the bias is comparable with (if not larger than) that obtained after separating the haloes based on the
half-mass formation time, $t_{50}$ \citep[see e.g.][]{Gao2005} although $t_{\rm c}$ correlates poorly with
$t_{50}$ (see also Sections~\ref{D-ZOMG} and \ref{sec_growth}). 

\begin{figure*}
 \includegraphics[width=0.3\textwidth,bb=0 0 348 360,keepaspectratio=true]{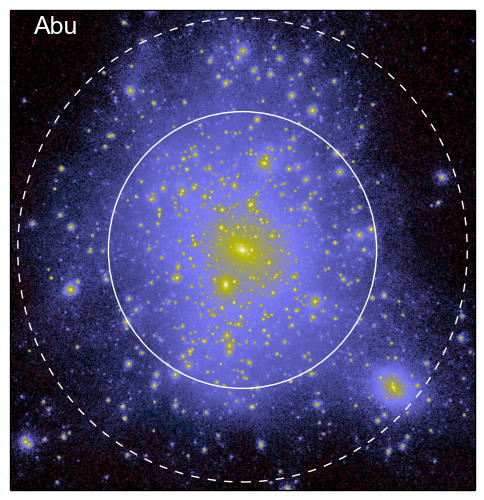}
 \includegraphics[width=0.3\textwidth,bb=0 0 348 360,keepaspectratio=true]{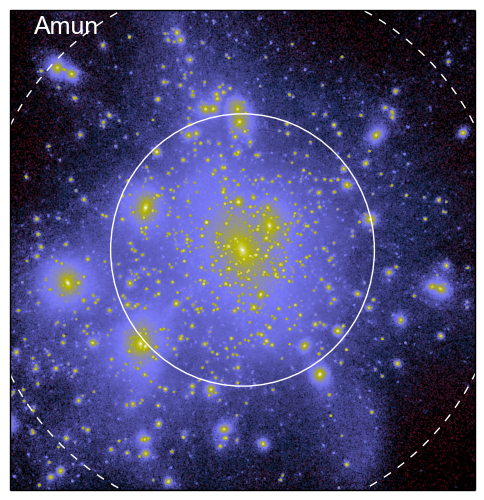}

 \hspace{\fill}
 \includegraphics[width=0.3\textwidth,bb=0 0 348 360,keepaspectratio=true]{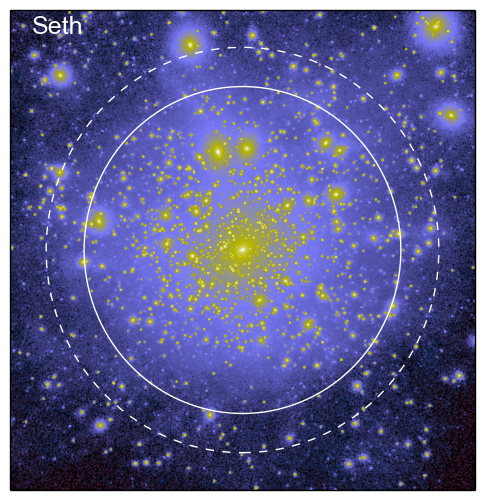}
 \hspace{\fill}
 \includegraphics[height=0.3\textwidth,bb=0 0 86 414,keepaspectratio=true]{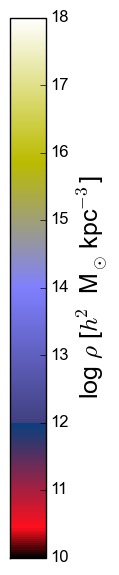}
 \hspace{\fill}
 \includegraphics[width=0.3\textwidth,bb=0 0 348 360,keepaspectratio=true]{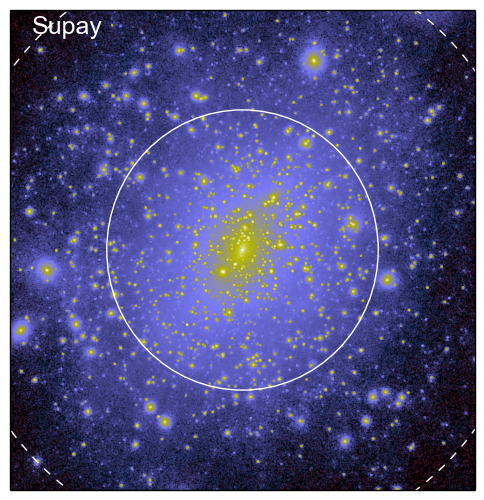}
 \hspace{\fill}

 \includegraphics[width=0.3\textwidth,bb=0 0 348 360,keepaspectratio=true]{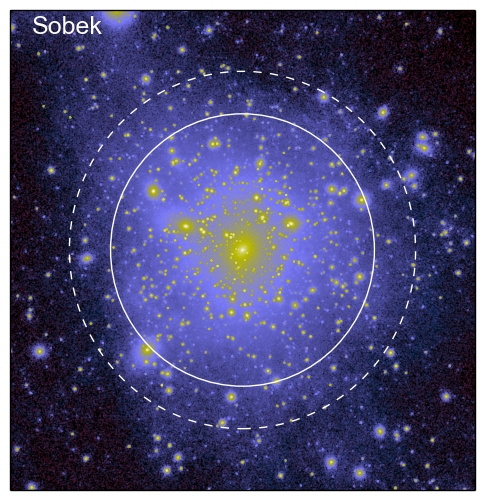}
 \includegraphics[width=0.3\textwidth,bb=0 0 348 360,keepaspectratio=true]{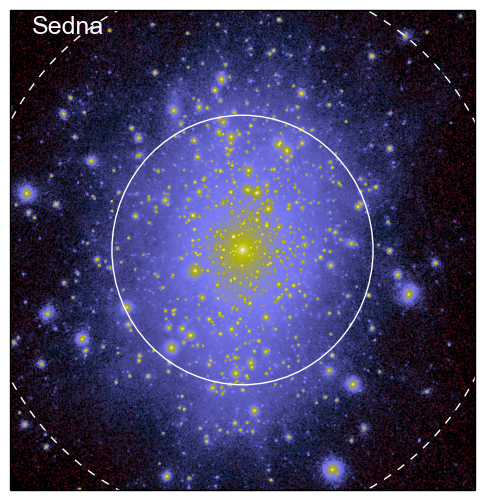}
 \includegraphics[width=0.3\textwidth,bb=0 0 348 360,keepaspectratio=true]{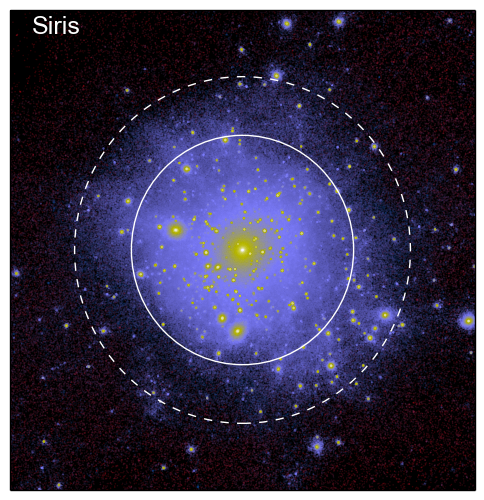}
 \caption{Matter distribution in the D-ZOMG haloes.
 $N$-body particles are coloured based on the local mass density computed with a standard smoothed-particle-hydrodynamic kernel including 64 neighbours.
 Each panel refers to a cube with side length of $540\,h^{-1}$ kpc projected along one of the axes and centred on a resimulated halo.
 The solid circle shows $R_{\rm h}$ while the dashed one indicates $R_{\rm spl}$.}
 \label{fig_haloes}
\end{figure*}

\begin{table*}
\caption{Properties of the haloes in the D-ZOMG simulation suite (Z4096).
The columns give the halo name and the colour we associate with it in all plots,
the $z=0$ mass ($M_{\rm h}$) and radius ($R_{\rm h}$), 
the splashback radius ($R_{\rm spl}$), 
the minimal Hill radius $R_{\rm H}$, 
the concentration ($c_{\rm NFW}$) of the radial mass-density profile fitted with the Navarro-Frenk-White (NFW) formula \citep*{NFW1997},
the major-to-minor axis ratio ($i_1/i_3$) derived from the inertia tensor of the halo particles in the initial conditions and at the present time as well as the angle between the shortest principal axes at these two epochs ($\theta_{33}$), the cosmic expansion factor at formation time defined in terms of the first appearance of a
progenitor with 30 and 50 per cent of the final halo mass ($a_{30}$ and $a_{50}$), the expansion factor at collapse time ($a_{\rm c}$)  defined as in Section~\ref{sec_age},
and, finally, the spin parameter ($\lambda_{\rm s}$).}
\begin{tabular}{ccccccccccccccc}\hline
 Halo & $M_{\rm h}$ & $R_{\rm h}$ & $R_{\rm spl}$ & $R_{\rm H}$ & $c_{\rm NFW}$ & $i_1/i_3$ & $i_1/i_3$ & $\cos(\theta_{33})$ & $a_{30}$ & $a_{50}$ & $a_{\rm c}$ & $\lambda_{\rm s}$ \\ 	
  & $(10^{11}h^{-1}{\rm M}_\odot)$ & $(h^{-1}$kpc) & $(h^{-1}$kpc) & $(h^{-1}$kpc) &  & $z=99$ & $z=0$ &  &  &  &  & $10^{-2}$ \\\hline						
\colorbox{ColorAbu}{\parbox{26pt}{\textcolor{white}{\bf Abu}}}		& 4.5 & 156 & 261 & 1528 & 8.0 & 1.39 & 1.23 & 0.71 & 0.44 & 0.46 & 0.97 & 2.4 \\		
\colorbox{ColorAmun}{\parbox{26pt}{\textcolor{white}{\bf Amun}}}	& 4.3 & 153 & 304 & 1697 & 9.3 & 1.58 & 1.56 & 0.90 & 0.50 & 0.51 & 1.00 & 3.4 \\		
\colorbox{ColorSeth}{\parbox{26pt}{\bf Seth}}				& 7.4 & 184 & 228 & 1478 & 6.0 & 3.45 & 1.10 & 0.70 & 0.32 & 0.49 & 0.39 & 3.7 \\		
\colorbox{ColorSupay}{\parbox{26pt}{\bf Supay}}				& 4.7 & 158 & 337 & 922 & 8.7 & 3.01 & 1.30 & 0.07 & 0.32 & 0.45 & 0.40 & 1.8 \\		
\colorbox{ColorSobek}{\parbox{26pt}{\textcolor{white}{\bf Sobek}}}	& 4.3 & 153 & 201 & 653 & 10.7 & 2.40 & 1.10 & 0.07 & 0.34 & 0.38 & 0.38 & 2.1 \\		
\colorbox{ColorSedna}{\parbox{26pt}{\bf Sedna}}				& 4.2 & 152 & 306 & 854 & 11.2 & 2.36 & 1.36 & 0.17 & 0.27 & 0.50 & 0.40 & 1.6 \\		
\colorbox{ColorSiris}{\parbox{26pt}{\textcolor{white}{\bf Siris}}}	& 2.6 & 129 & 195 & 724 & 15.4 & 1.79 & 1.16 & 0.35 & 0.25 & 0.45 & 0.33 & 0.7 \\\hline	
\end{tabular}
\label{tab_coll_a}
\end{table*}

\subsection{The Dark-ZOMG simulation suite}
\label{D-ZOMG}
We employ high-resolution zoom simulations to pin down the physical mechanism leading to the different collapse histories presented in Fig.~\ref{fig_volume_example}.
At the level of single objects, the zoom technique provides several advantages over standard cosmological simulations in which galaxy-sized haloes are typically resolved with only a few thousand computational elements. Zooming allows a much closer look at the geometry of matter accretion and provides a detailed picture of the phase-space structure. 
In particular, we use this information to make a connection between the halo environment and its internal structure.
On the other hand,  the heavier computational requirements of zoom simulations limit our analysis to a small number of representative cases. The concept is to provide clear templates
and extract regularities from them that serve as a model to explain what happens in general.
Our investigation is thus complementary to statistical studies analysing thousands of low-resolution haloes.
In particular, we build upon \citet{Hahnetal2009} and \citetalias{Borzyszkowski2014}.

Target haloes for re-simulation are selected from the U512 run based on $M_{\rm h}$, $t_{\rm c}$, and applying an isolation criterion.
We consider haloes with mass $M_{\rm h}\simeq {\rm a\ few} \times10^{11} h^{-1} {\rm M}_\odot$ \citep[i.e. the hosts of present-day $L_*$ galaxies,][]{Mosteretal, Behroozi} that are not closely surrounded by more massive neighbours (i.e. no particles belonging to a more massive halo must be found within three $R_{\rm h}$ from the halo centre). 
The latter condition is necessary to limit the size of the high-resolution region and thus the computational costs while fully benefiting from the zoom-in method.
Admittedly, introducing the seclusion criterion might bias our sample. However, there are good reasons why the bias should be very small.
In the parent run, only 4.8 per cent of the haloes with the selected mass violate our isolation criterion.
Moreover, it has been shown that close encounters and fly-bys cannot generate the assembly bias  \citep{Wang2007, Hahnetal2009, Wangetal2011}.
Therefore, we are confident that our sample is adequate for this study and mostly representative of the halo population.

Among the haloes that satisfy the selection criteria,
we randomly pick five objects that collapsed very early on (at $a_{\rm c}\leq 0.4$) and two that have not collapsed by today.
This set forms the Dark-ZOMG simulation suite (D-ZOMG).
To facilitate distinguishing between the different haloes, we name each of them after an ancient god (and also associate them with individual colours in figures, see Table~\ref{tab_coll_a}).
We use names starting with the letter A (standing for `accreting') for the haloes with $t_{\rm c}\gtrsim t_0$ and with the letter S (standing for `stalled') for the remaining ones. 
Although the meaning of our labels can be intuited from Fig.~\ref{fig_volume_example}, it will become clearer in Section~\ref{sec_growth}.

For each halo,
we use the {\sc music} code \citep{HahnAbel11}  to generate zoom initial conditions in which
the mass resolution varies within the simulation box.
The high-resolution region coincides with the Lagrangian convex hull of the particles that in the parent simulation are found within $3R_{\rm h}$ from the halo centre of mass at $z=0$. 
Within this patch, matter is sampled with a large number of elements and, during the evolution, gravitational forces are computed at high resolution. Outside this volume, we place a number of buffer regions in which the simulation particles become progressively more massive (and forces less accurate) with increasing distance from the high-resolution region.
This method allows us to simulate the haloes and their immediate surroundings at high resolution while keeping into account the tidal influence of the large-scale structure in a reasonable computational time.
In the D-ZOMG simulations,
the coarsest mass resolution in the box corresponds to that of a uniform grid sampled with $128^3$ particles
while we simulate each halo twice at different maximum resolutions. In our production runs (Z4096), the dark matter particles
in the high-resolution region have a mass of $1.6\times10^{5}~h^{-1}{\rm M}_\odot$ corresponding to
a uniform grid with $4096^3$ elements. In order to test the robustness of our results with respect to 
numerical errors, we repeat the simulations using a particle mass of $1.3\times10^{6}~h^{-1}{\rm M}_\odot$ 
corresponding to a $2048^3$ grid (Z2048).
The Plummer equivalent softening lengths adopted in the high-resolution region 
are $0.24$ and $0.49\,h^{-1}$kpc for Z4096 and Z2048, respectively.

During each run, we save 8 snapshots between $z=90$ and $z=30$ uniformly sampled in redshift,
and 12 between $z=30$ and $z=9$. Subsequently, we  
save an output every 20 Myr, for a total of 682 snapshots.
We identify haloes in all snapshots (using the method previously described) and find their main progenitor by maximizing the merit function
$N_{i\cap j}^2/(N_iN_j)$ where $N_i$ and $N_j$ denote the number of particles in the descendant halo
and in the candidate progenitor halo (defined at the previous snapshot with respect to the descendant), respectively, and $N_{i\cap j}$ is the number of particles they have in common \citep{KnollmannKnebe09}.
Iterating this procedure, we build mass-accretion histories moving along the so-called `main branch' of the halo merger tree \citep[e.g.][]{LiMoGao2008}.
We tag the earliest time at which a main-branch progenitor with a mass of $M_{\rm h}/2$ appears as $t_{50}$
and we refer to this quantity as the half-mass formation time \citep{LaceyCole93}.
Similarly, we use the symbol $t_{30}$ to indicate the time at which a progenitor with 30 per cent of the final
mass first comes into existence.

A particular, parameter-free, halo definition has been recently discussed in the literature (\citealt{Diemeretal2014,Adhikarietal2014}; \citealt*{Moreetal2015}). 
In this case, the halo boundary is defined in terms of the so-called `splashback' radius, $R_{\rm spl}$, which corresponds to
a sudden steepening of the radial mass-density profile and is attributed to a caustic (intended as the accumulation of particle trajectories) located near the first apocentre
of recently accreted matter.
We determine the splashback radius for our re-simulated haloes (at all time steps) by measuring the density profile in spherical shells around the halo centre and locating the global minimum of its slope.
Since substructures introduce spikes in the profile, we preventively exclude their particles from the calculation.
[Substructures are associated with local density peaks within haloes and their edge coincides with the radius
of the smallest sphere enclosing all their gravitationally bound particles.]
The mass enclosed within the splashback radius is denoted by $M_{\rm spl}$ and we track its growth 
along the main branch of the merger tree. 

Before moving further with our study,
it is imperative to test that our definition of $t_{\rm c}$ is robust with respect to numerical resolution.
Fig.~\ref{fig_volume_example} shows that the U256, U512, Z2048 and Z4096 simulations give perfectly
consistent values for the collapse times of Amun and Supay. The same conclusion applies to the other 
re-simulated haloes (not shown in the figure). The only discrepancy which is worth mentioning is found
for Abu. In this case, we measure $a_{\rm c}>1$ in the U256 and U512 (Fig.~\ref{fig_colltime_dist}) runs as well as $a_{\rm c}=0.96$ and 0.97 in Z2048 and Z4096, respectively (Table~\ref{tab_coll_a}). 
The difference is caused by a merger with a sizeable mass ratio that takes place at $a\sim 0.9$. The outward motion of the recently accreted subhalo after the first pericentre passage causes a small upturn in the $V(t)$ plot (we remind the reader that all our ellipsoids are drawn around the centre of mass) which basically determines the collapse time. 
In fact, the integral in equation~(\ref{eqdeftc}) is sensitive to fluctuations in $V(t)$ when $t_{\rm c}\simeq t_{\rm max}$ while it is very robust when $t_{\rm c}\ll t_{\rm max}$ so that the interval of integration is larger. The net effect is that small temporary increases in $V(t)$ at late times can artificially give values of $t_{\rm c}$ which are very close to $t_{\rm max}$ in high-resolution runs. This issue is irrelevant if one wants to distinguish early and late collapsing haloes 
and can be easily solved by using a larger value for $t_{\rm max}$ (i.e. by running the simulations longer).
Inspection of the mass shells immediately surrounding Abu reveals that they are coherently infalling at $z=0$ as expected for an accreting halo.
It follows from this test that large-volume cosmological simulations
with uniform sampling of the initial conditions are well suited for measuring $t_{\rm c}$ and performing statistical studies based on this quantity.

\section{Results}\label{sec_results}
A synopsis of the D-ZOMG haloes is given in Fig.~\ref{fig_haloes} and Table~\ref{tab_coll_a} where we report several halo properties measured from the Z4096 simulations. A prominent feature can be noticed at first glance:
stalled haloes originate from much more aspherical patches in the initial conditions than the
accreting ones (while the final shape of the haloes does not seem to be different
between the two classes). In the remainder of this paper we will investigate the origin of this characteristic pattern and delve into the physical processes that generate it.

\subsection{Collapse time and halo growth}\label{sec_growth}
At this point, it is interesting to compare our definition of collapse time with other related quantities that are commonly used in the literature to define halo ages. 
From Table~\ref{tab_coll_a} we note that,
while accreting and stalled haloes neatly separate based on $t_{\rm c}$, they are not
so clearly distinct using the half-mass formation time. As a matter of fact, many of our re-simulated haloes have very similar formation times. On the other hand, using $t_{30}$ 
gives results more in line with those obtained with $t_{\rm c}$. This suggests that $t_{\rm c}$ contains
information on the early phases of halo formation.

\begin{figure}
 \includegraphics[width=0.5\textwidth,bb=0 0 699 518,keepaspectratio=true]{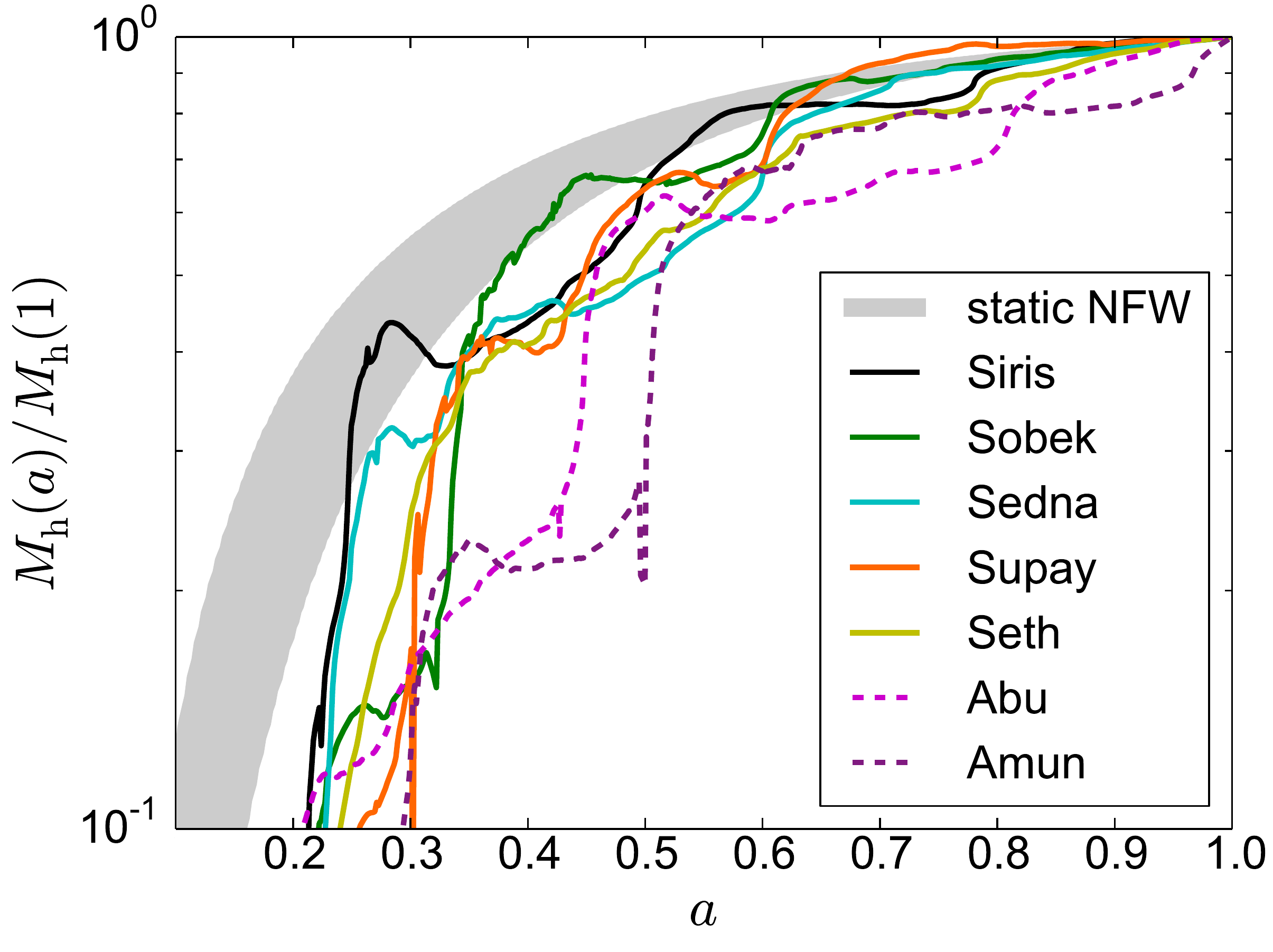}
 \caption{Mass accretion histories of the resimulated haloes in the highest resolution simulation.
 The grey shaded area shows the pseudo-evolution of a static NFW profile due to the changing background density in the range of concentrations of our simulated haloes $6<c_{\rm NFW}<16$.
 See text for details.}
 \label{fig_mah_resim}
\end{figure}

The mass-accretion histories of our re-simulated haloes are shown in Fig.~\ref{fig_mah_resim}.
Abu and Amun experience a major merger at $a\sim 0.4$-0.5, steadily grow in mass afterwards 
and undergo a minor merger at late times. In parallel,
the stalled haloes assemble a significant fraction of their mass early on through major mergers
and subsequently show a positive net mass-increment rate.
From this perspective, it appears that the difference between stalled and accreting haloes is just the timing at which an adequately sized progenitor is assembled. No other evident dissimilarity emerges from Fig. ~\ref{fig_mah_resim}.
We are going to show that this mental picture is simplistic and does not capture a number of 
fundamental features that distinguish the two classes of haloes.

\begin{figure}
 \includegraphics[width=0.49\textwidth,bb=0 0 696 514,keepaspectratio=true]{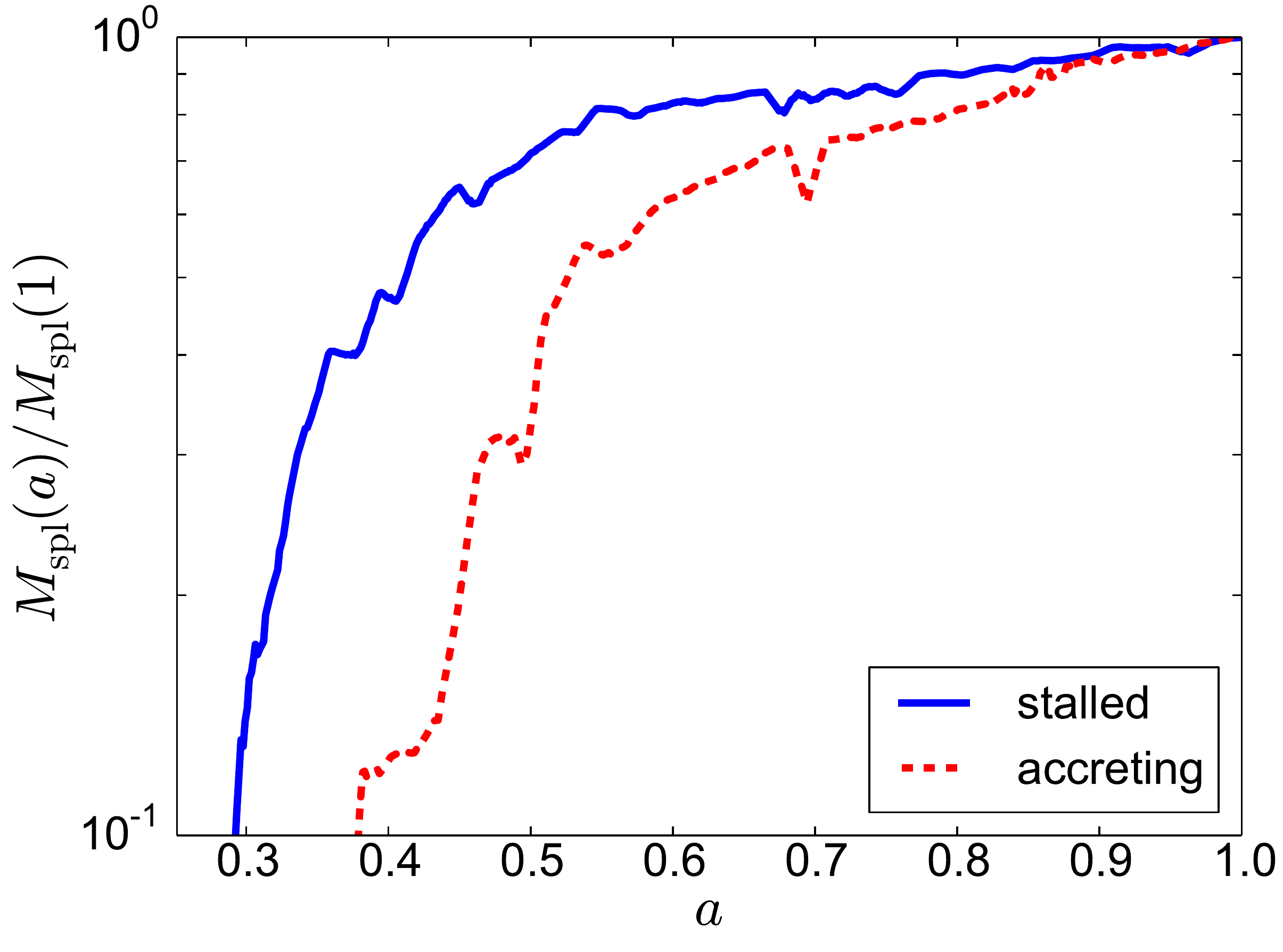}
 \caption{The mass enclosed within the splashback radius is plotted as a function of the cosmic expansion factor.
Haloes are followed along the main branch of their merger tree.
To reduce noise, the curves have been obtained 
averaging over our accreting (dashed red) and stalled (solid blue) haloes.
The time evolution has also been coarse-grained with a resolution of $200$ Myr.
Stalled haloes assemble most of their final mass early on compared to accreting haloes which, on the other hand, grow more rapidly at the present time.}
\label{fig_splashback}
\end{figure}

The first thing to consider is that
haloes are conventionally cut out of their environment adopting a somewhat arbitrary criterion that defines their outer boundary. 
Following a standard practice, 
we use a density threshold which decreases with time.
As a consequence of this choice, the halo mass is subject to pseudo-evolution, meaning that the very same immutable object would be assigned different masses at different epochs just because its conventional boundary moves outwards \citep*[see e.g.][]{Diemeretal2013,Zemp2014}.
The grey band in Fig.~\ref{fig_mah_resim} highlights the pseudo-evolution of a stationary NFW density profile
and its thickness reflects the range of concentrations covered by our
stalled haloes ($6<c_{\rm NFW}<16$). For $a>0.6$, the mass-accretion histories of Sedna, Sobek and Supay 
closely match the pseudo-evolution of a stationary NFW profile while this is not true for the accreting haloes.
In fact, Abu and Amun grow faster than that mainly because they both undergo a merger with approximately 1:6 mass ratio.
Finally, Seth and Siris show an intermediate behaviour as they also experience a merging event at $a\sim 0.77$ (with a substantially smaller mass ratio) and
follow pseudo-evolution afterwards.

\begin{figure*}
 \includegraphics[width=0.49\textwidth,bb=0 0 697 567,keepaspectratio=true]{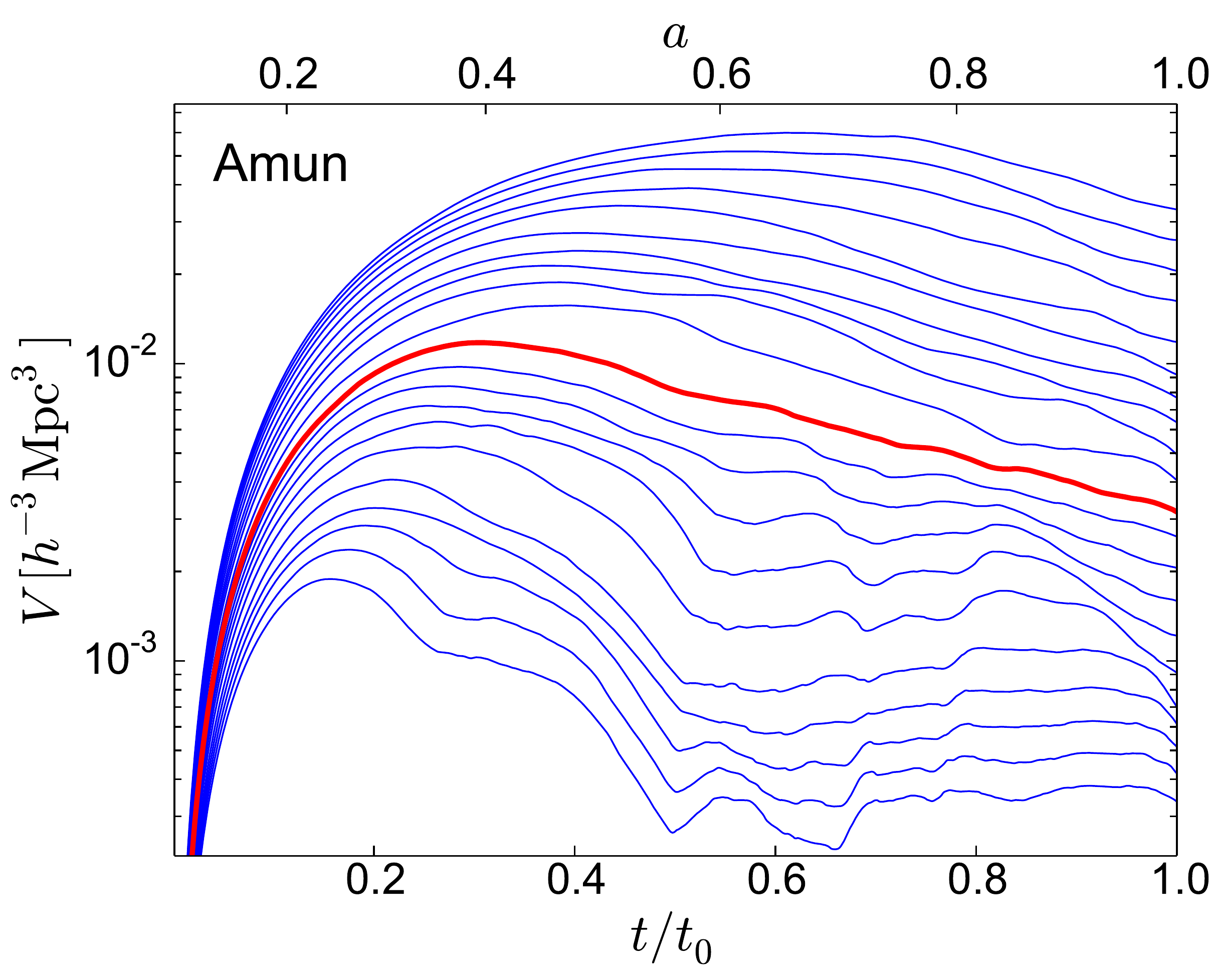}
 \includegraphics[width=0.49\textwidth,bb=0 0 697 567,keepaspectratio=true]{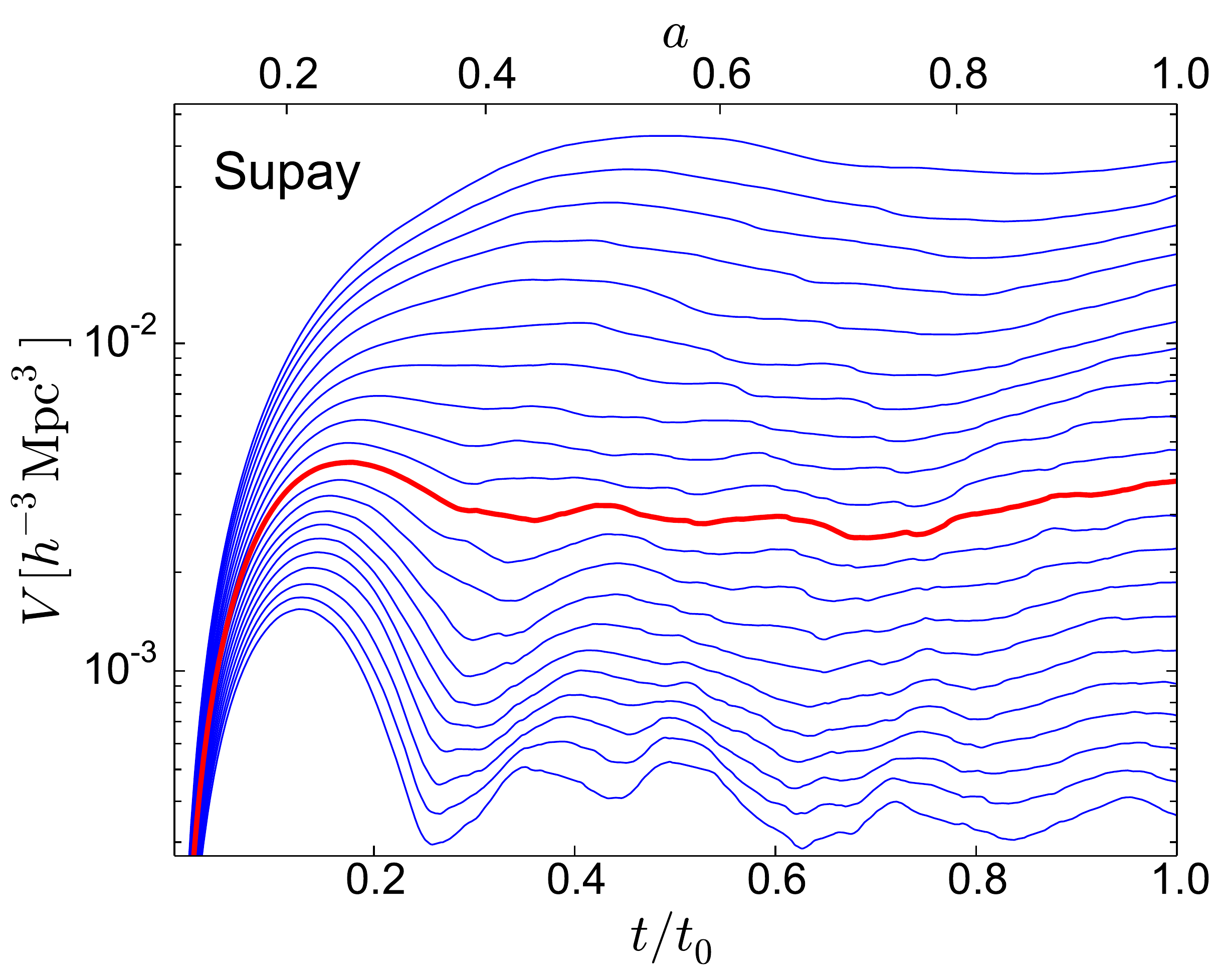}
 \caption{The red curves reproduce the volume evolution shown in Fig. 1 for Amun (left-hand panel) and Supay (right-hand panel).
The same analysis is repeated considering different volumes (blue curves) that enclose a mass ranging from $0.1 M_{\rm h}$ to $10 M_{\rm h}$ in logarithmic steps.
To improve readability and place equal emphasis on all shells, a logarithmic scale is used for $V$.
Note that the outer mass shells are infalling on to Amun and receding from Supay.}
\label{fig_volume_multishell}
\end{figure*}

In Fig.~\ref{fig_splashback} we show the evolution of $M_{\rm spl}$ which is not affected by changes in
the critical density of the universe. Since measurements of $M_{\rm spl}$ are quite noisy for individual haloes, we report results averaged over our re-simulated objects and smoothed over 10 snapshots. 
These curves are well fitted by an exponential function of redshift \citep{Wechsleretal2002} only for $z<1$. 
At earlier times, the evolution is much more rapid than the simple analytical approximation.
Two things are worth noticing here. (i) The haloes with low $t_{\rm c}$ assemble any given fraction 
of their final mass earlier with respect to the others. (ii) These haloes do not grow much at late times:
on average, $M_{\rm spl}$ only increases by 17 per cent since $a=0.6$.
On the other hand, the haloes with high $t_{\rm c}$ increase their mass by 34 per cent during the same period. 
All this points to $t_{\rm c}$ as a discriminator between haloes that assemble at early times 
and then undergo little changes and those that gathered most of their mass later on and are still growing at the present time.

\subsection{Collapse time and accretion dynamics}
Although parameter free, our definition of collapse time is sensitive to the halo definition.
In fact, the halo finder identifies the particle set and thus the Lagrangian patch whose volume evolution determines $t_{\rm c}$. 
Let us imagine for a moment that our halo definition just picks the innermost virialized core of the halo and excludes more recently accreted shells 
[haloes are known to form from the inside out (\citealt*{Salvador1998}; \citealt{Wechsleretal2002,Zhaoetal2003,Wangetal2011})].
In this case, we would severely underestimate the collapse time for the more extended halo.
To investigate the stability of $t_{\rm c}$ with respect to the operational halo definition, 
we repeat the study of the volume evolution for many shells that enclose
from 10 per cent up to 10 times the final halo mass (in logarithmic steps).
Our results for Amun and Supay are plotted in Fig.~\ref{fig_volume_multishell} (we find consistent
results also for all the other re-simulated haloes).
It is straightforward to note that the innermost core of Amun reaches a stable configuration at approximately
$t=t_0/2$ while all the outer shells are still contracting towards the halo centre at the present time
(hence the appellative `accreting').
This conforms to the common idea that haloes constantly grow in mass due to the infall of outer shells
\citep[e.g.][]{Press-Schechter,BCEK}. In this case, it would make sense to define the boundary of the halo
as the location of the outermost shell that stabilized by $z=0$. For Amun, this criterion gives a
mass of $\sim0.5\,M_{\rm h}$.
Switching our attention to Supay, we immediately note that there are no infalling shells at the
present time. The material surrounding what we conventionally call the halo is actually receding from its centre.
Even some of the internal shells show the same feature. 
The net outward motion kicks in at  $t> 0.8\, t_0$ following a long period in which the shells have been loitering 
at nearly constant volume. Based on this information it is unclear how to draw a physically motivated boundary 
for Supay. Anyway, our analysis suggests that Supay, as a physical object, should not be able to grow in mass
at late times (hence the appellative `stalled').

\begin{figure*}
 \includegraphics[width=0.9\textwidth,bb=0 0 1128 687,keepaspectratio=true]{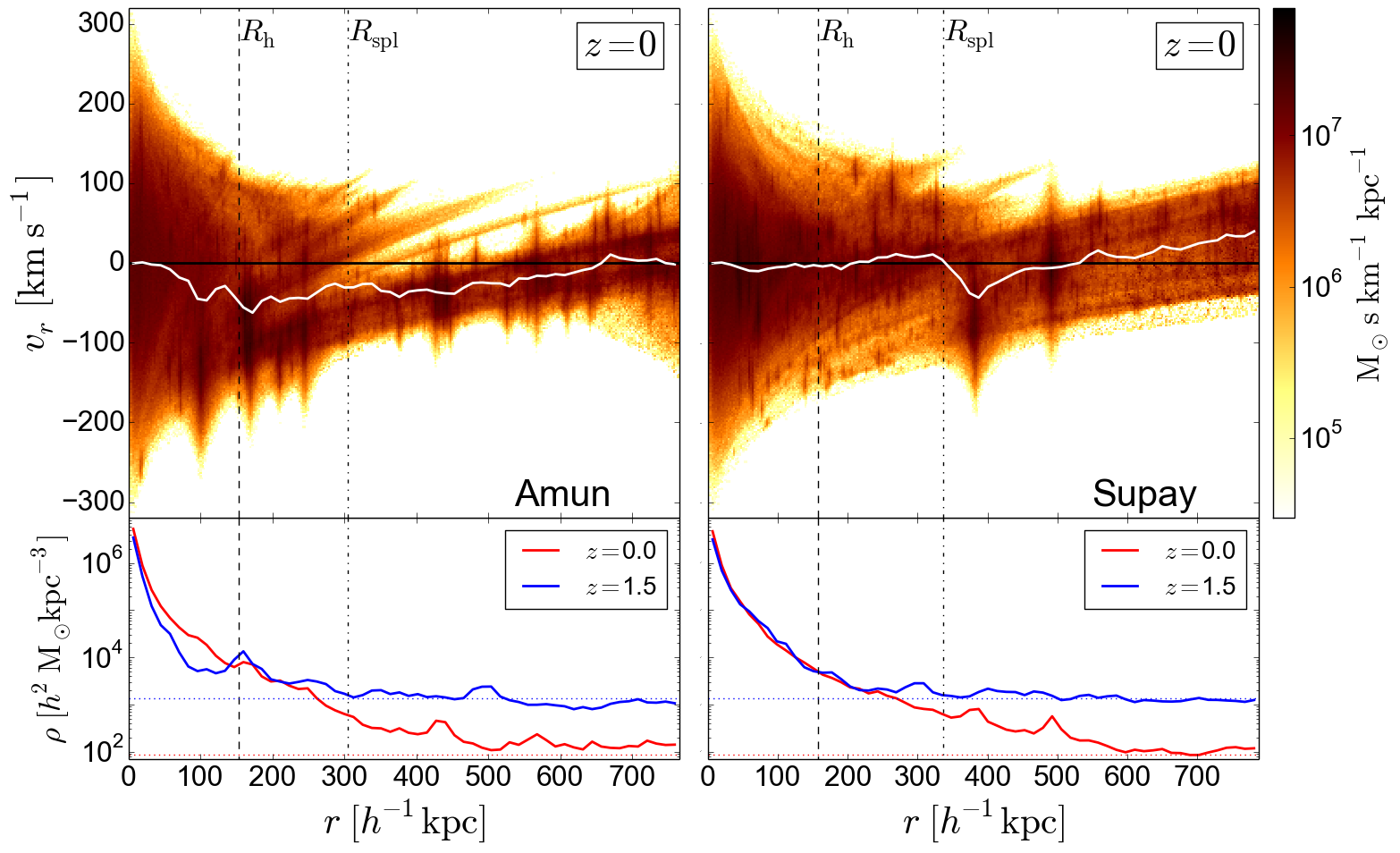}
 \caption{Top: radial phase-space diagram for Amun (left-hand panel) and Supay (right-hand panel) at $z=0$.
The colour coding indicates the matter density in phase space while the white curve shows the mean velocity
as a function of the distance from the halo centre.
The vertical lines refer to the halo radius (dashed) and the splashback radius (dot-dashed).
Bottom: the radial mass density profiles of the haloes at $z=0$ (red) and 1.5 (blue).
The dotted horizontal lines indicate the mean matter density of the universe at the corresponding epochs.}
\label{fig_phase_space}
\end{figure*}

Haloes are objects with a complex dynamics and continuously interact with their surroundings. 
To capture the details that might have been lost in the study of the different mass shells,
in the top-panels of Fig.~\ref{fig_phase_space} we investigate the radial phase-space distribution of matter at $z=0$ within and around Amun and Supay (consistent results are found for all the other re-simulated haloes).
Colour coding reflects the phase-space density while the vertical lines indicate $R_{\rm h}$ (dashed)
and $R_{\rm spl}$ (dot-dashed).  
The halo cores are characterized by a large velocity dispersion reflecting the depth of the gravitational-potential well. Substructures and external haloes can also be easily identified as patches with high density and increased
velocity dispersion. Infalling material ($v_r<0$) moves faster and faster while approaching the halo core and,
after reaching the pericentre, suddenly reverses its motion ($v_r>0$).
Particles that are bound to the halo describe particular trajectories in the radial phase-space diagram.
The needle-shaped configurations with positive radial velocity correspond to the remnants of substructures 
after their first pericentre passage. Note that these streams of matter stretch far beyond $R_{\rm spl}$ which has been computed in terms of a spherically averaged mass-density profile. Defining the halo boundary
in terms of the apocentre of the material `ejected' with high speed would thus require tracing the positions
of the caustics in three dimensions.
The thick white curve in Fig.~\ref{fig_phase_space} shows the mean radial velocity as a function of the distance, $r$, from the halo centre. Amun shows a net radial inflow of matter over a large range of $r$. This flow is rather cold and coherent.
Basically the phase-space diagram can be decomposed into three elements: the halo, an accretion flow and
the outward streams (although it is not obvious how to separate them).
The corresponding diagram for Supay shows striking differences. For instance, the outward streams are much less pronounced than in Amun. Moreover, for Supay there is no net inflow of matter:
the mean radial velocity is zero in the innermost regions and positive in the outer regions.
This does not mean that no material is infalling, rather that the mass outflow through any spherical surface
exactly matches the inflow up to distances of several times $R_{\rm h}$ \citep[consistent with ][]{Pradaetal2006}.
All this suggests that 
some sort of dynamical equilibrium is established between the halo and the surrounding environment which is characterized by a rather large velocity dispersion. 

The bottom panels of Fig.~\ref{fig_phase_space} show the mass-density profile of the two haloes at $z=0$ and 
of their progenitors at $z=1.5$ (which corresponds to the collapse time of Supay).
The matter distribution in Amun has evolved a lot between these two epochs. Only the very central and dense core appears not to have changed while the outskirts have developed later. 
On the contrary, Supay shows exactly the same profile at $z=1.5$ and at $z=0$ within the innermost $250\,h^{-1}$ kpc ($\simeq1.6 \,R_{\rm h}$) while the evolution in the
outer regions just reflects the drop in the mean density of the universe. 

We have now collected sufficient factual evidence to support the claim that there exists a large population of dark matter haloes that
(i) are in place since early times with essentially unchanged structure;
(ii) do not grow in mass although they experience some matter infall which is however balanced by
outflows.
These stalled haloes form the bulk of galaxy-sized haloes at the present time (Fig.~\ref{fig_colltime_dist}).
Our findings challenge the common wisdom about halo formation according to which the halo mass constantly grows with time due to infalling material. 
Only a small minority of the haloes show this behaviour on galaxy scales. 
Conversely, the vast majority of cluster-sized haloes are of the accreting type \citepalias{Borzyszkowski2014}.
What remains to be identified is the physical process that prevents accretion to the stalled haloes. This is the subject of the following section.

\begin{figure*}
\centering
\includegraphics[width=0.64\textwidth,bb=0 0 864 432]{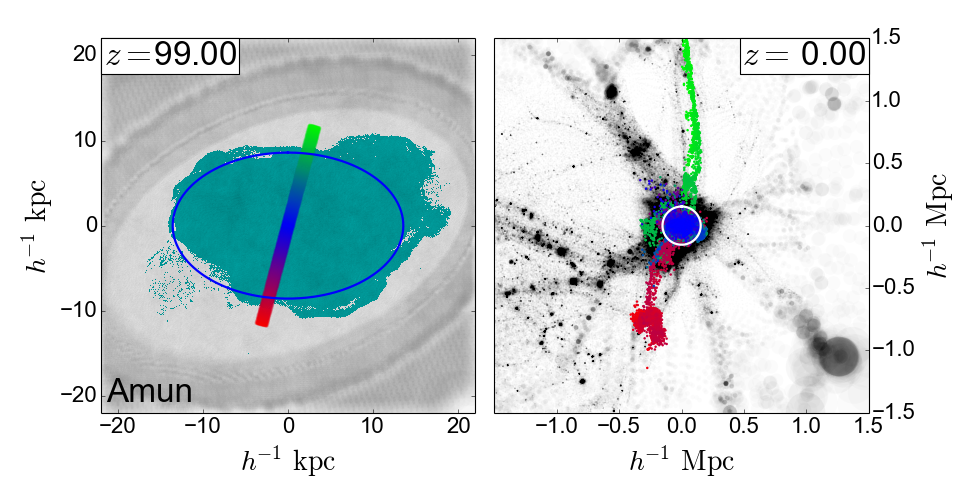}
\hspace{\fill}
\includegraphics[width=0.32\textwidth,bb=0 0 432 432,keepaspectratio=true]{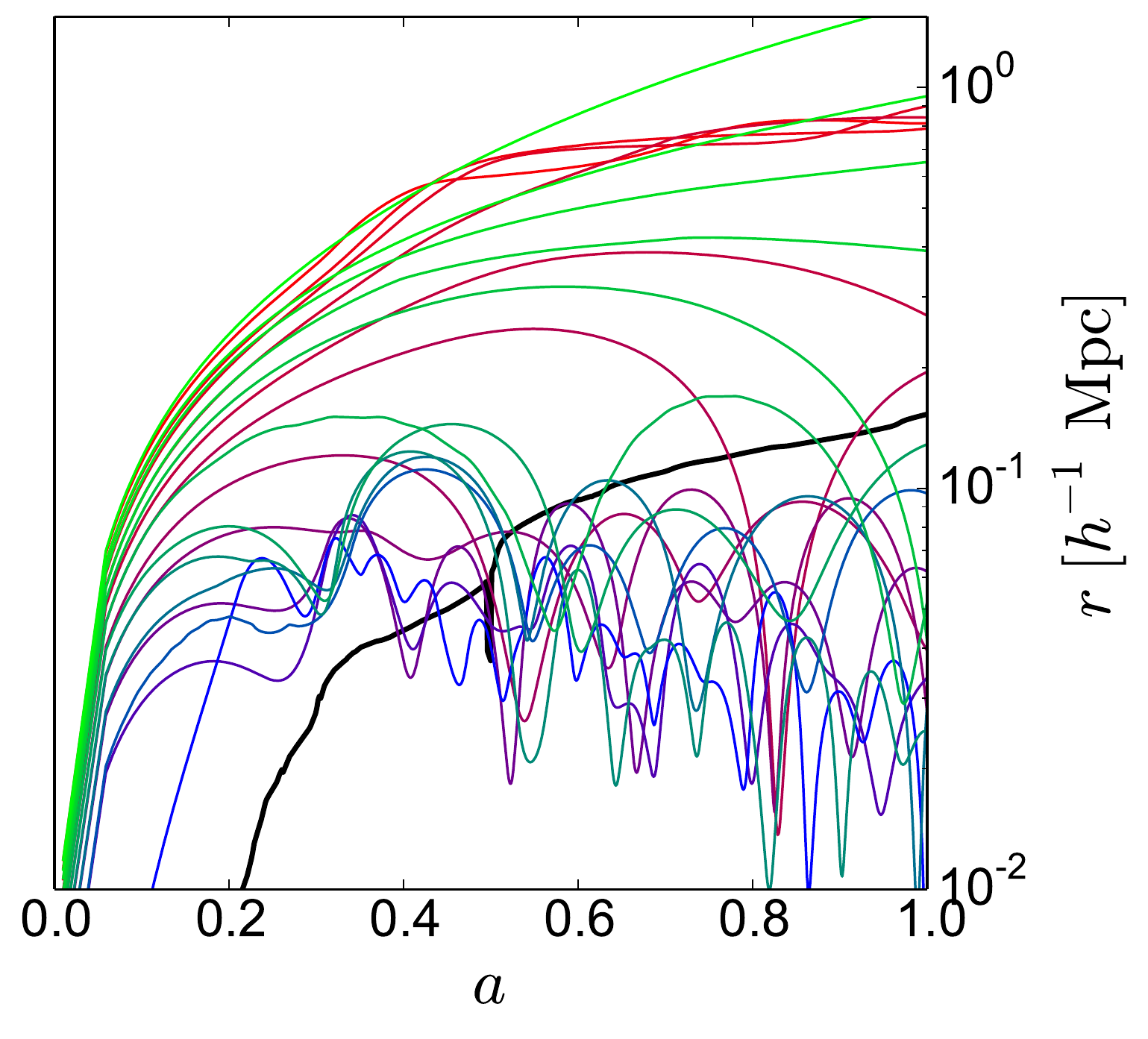}
\includegraphics[width=0.64\textwidth,bb=0 0 864 432]{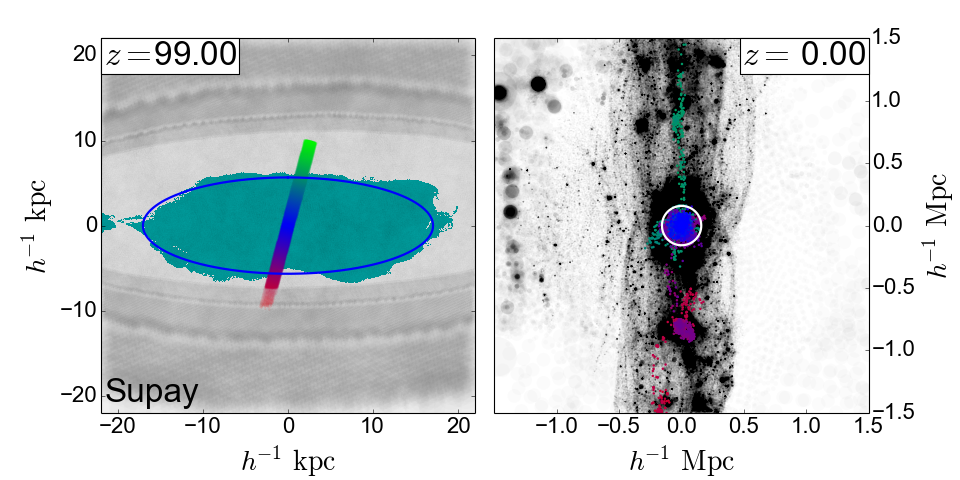}
\hspace{\fill}
\includegraphics[width=0.32\textwidth,bb=0 0 432 432,keepaspectratio=true]{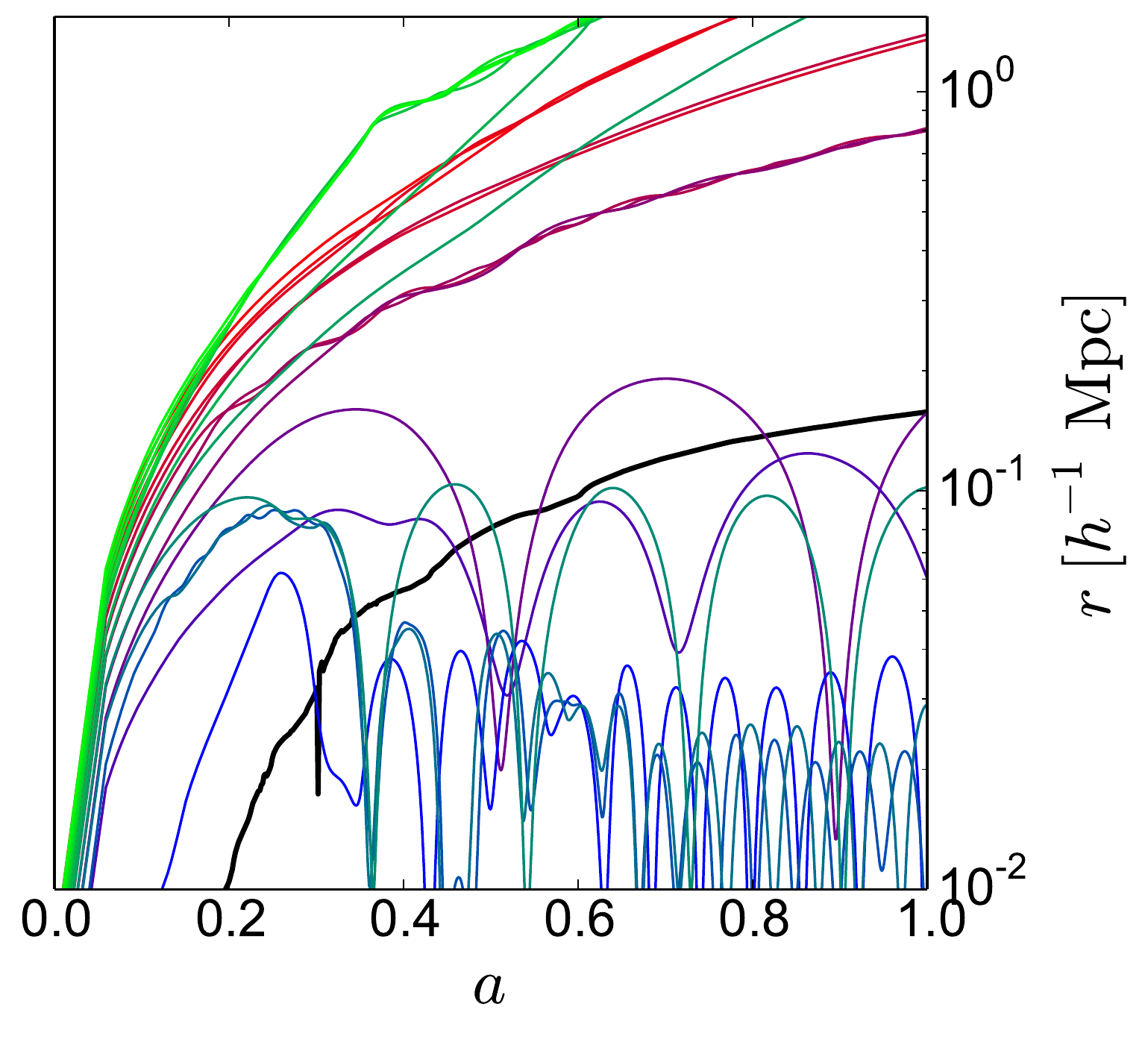}
\includegraphics[width=0.64\textwidth,bb=0 0 864 432]{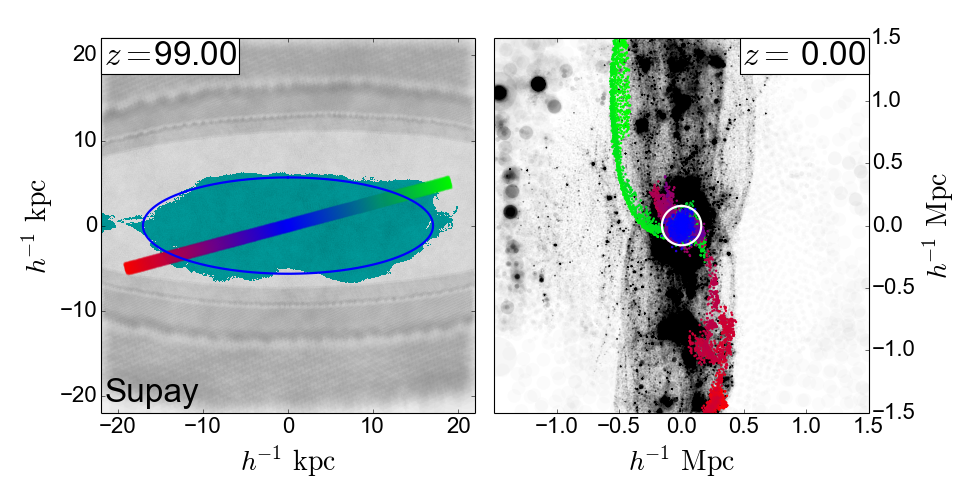}
\hspace{\fill}
\includegraphics[width=0.32\textwidth,bb=0 0 432 432,keepaspectratio=true]{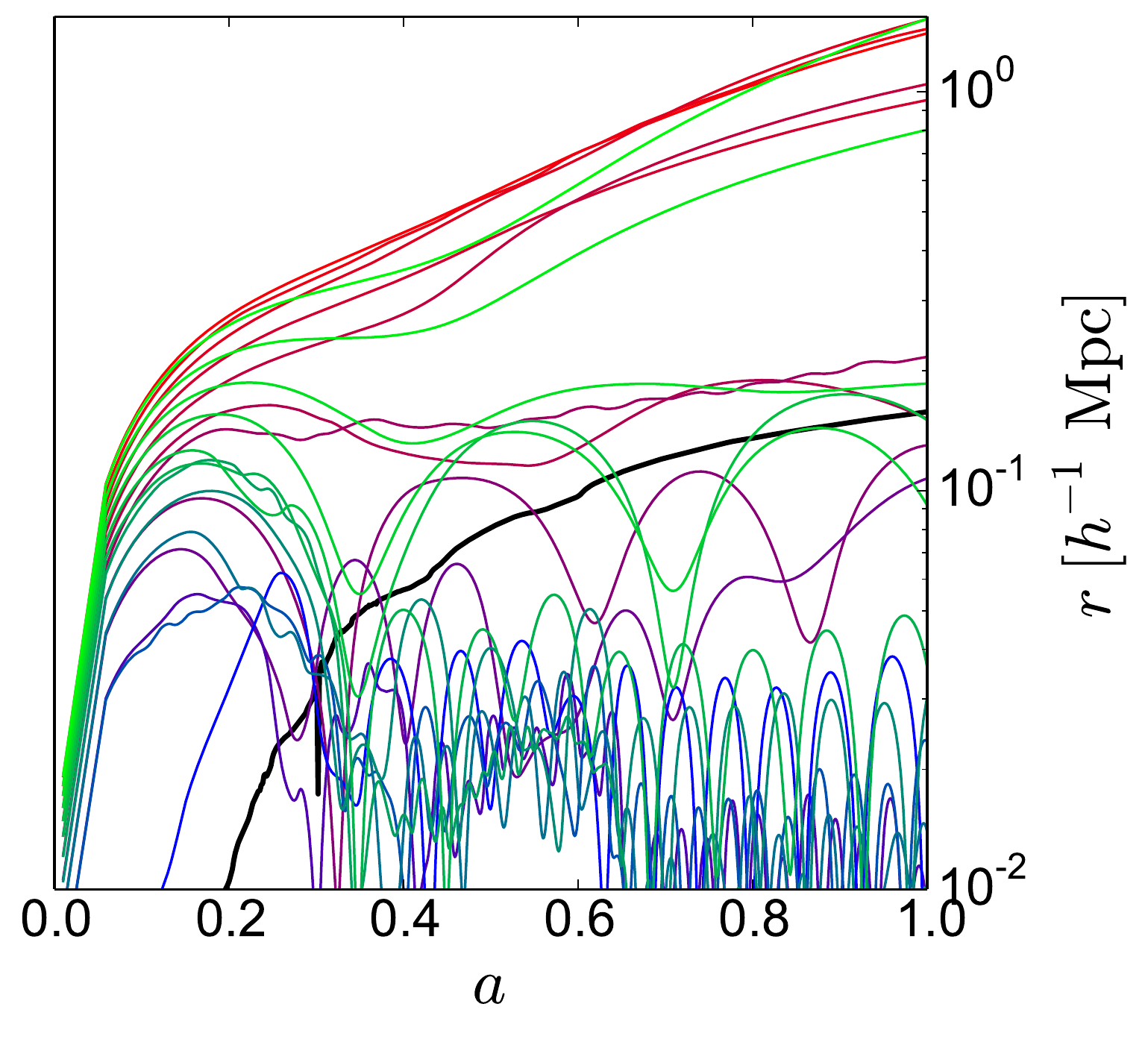}
\caption{Particle positions and orbits for Amun (top) and Supay (middle and bottom).
The left-hand panels show the halo particles in the initial conditions (cyan points). 
The $x$ and $y$ axes are oriented parallel to the longest and the shortest principal axes of inertia of the halo particles at $z=99$, respectively.
The blue ellipse shows the projection of the inertia ellipsoid on this plane.
The underlying grey-scale reveals the multi-mass sampling of the initial conditions. 
The central light grey area denotes the high-resolution region.
We select particles within a thin cylinder and colour them according to their distance from the centre of mass of the protohalo.
The projection of the cylinder on to the plane of the figure produces the central coloured rectangle.
The middle panels show the particle distribution at $z=0$ using the same orientation of the axes as in the left-hand panels (the size of the represented region is different though).
All the particles within $\pm 100\,h^{-1}$ kpc from the plane of the page are rendered using grey circles with size proportional to the particle mass and fixed transparency. 
Note that the middle and bottom row only differ by the orientation of the thin cylinder used to define the coloured particles.
The white circumference indicates the conventional halo boundary, $R_{\rm h}$.
The final positions of the particles that initially lie within the cylinder shown in the left-hand panel are highlighted using the colour coding introduced above.
The right-hand panels show how the distance to the halo centre of 21 particles selected within the cylinder evolves with time (coloured curves). 
Also indicated is the evolution of the halo radius (thicker black curve).
The halo position and size are computed considering the main progenitor at every snapshot.}
\label{fig_part_dist}
\end{figure*}

\begin{figure*}
 \includegraphics[width=0.9\textwidth,bb=0 0 1128 529,keepaspectratio=true]{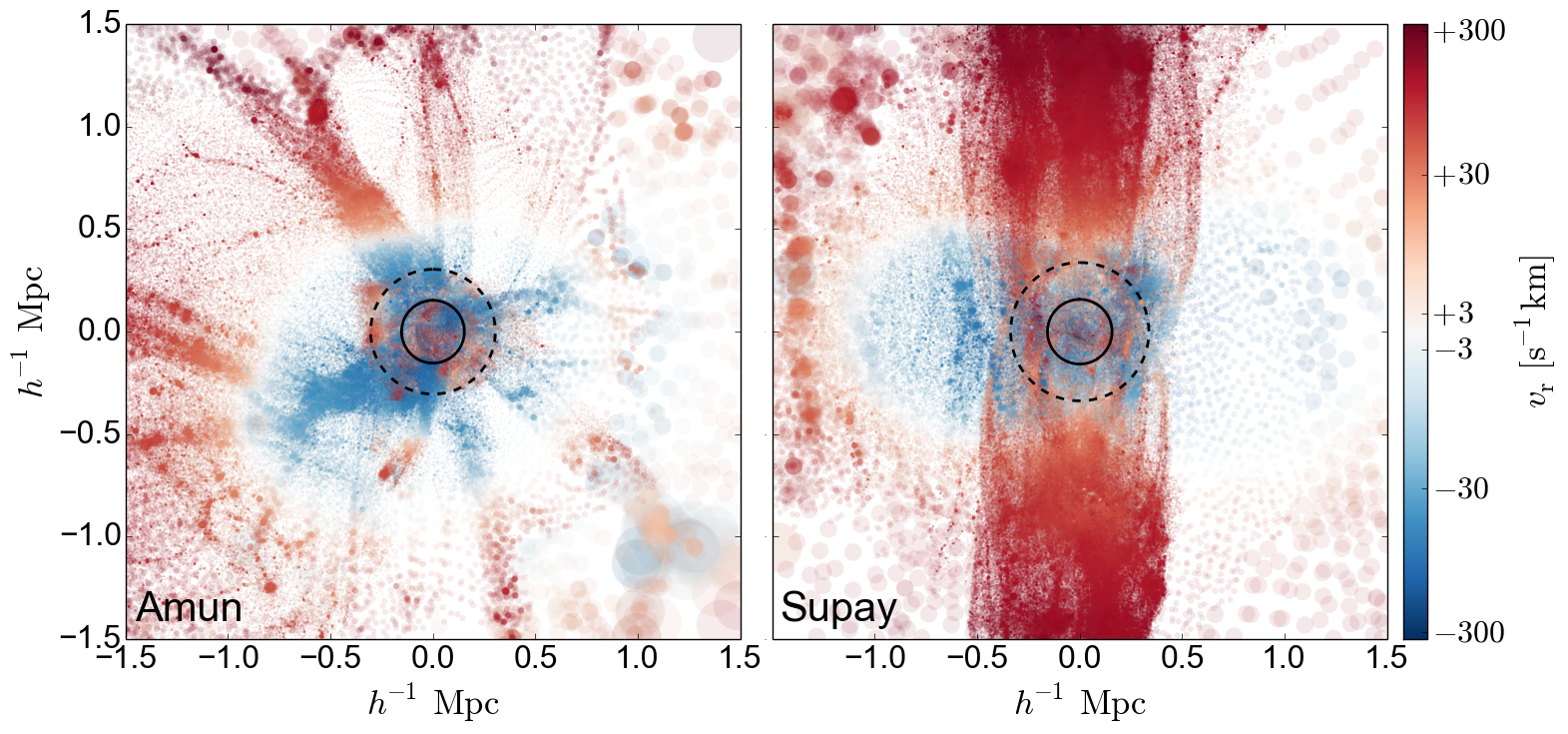}
 \caption{As in the middle panels of Fig.~\ref{fig_part_dist} but colouring the simulation particles according to their radial velocity with respect to the halo centre. 
 Note that separate logarithmic scales are used for infalling and receding matter connected with a linear scale in the range from $-3$ to $+3\,{\rm km\,s^{-1}}$. 
 The solid and dashed circles indicate the halo and the splashback radius, respectively.}
\label{fig_halo_acc}
\end{figure*}

\subsection{The impact of the cosmic web on halo growth}\label{sec_envi} 

We want to study how accreting and stalled haloes are assembled and connect this process with the
properties of their Lagrangian patches and final environments. To achieve this goal, we use Amun and Supay as templates representing the two classes of haloes [the other haloes behave identically]. 
The cyan area in the left-hand panels of Fig.~\ref{fig_part_dist} marks the distribution of the halo particles 
in the initial conditions that from now on we call the `protohalo'.
We use physical coordinates and project the particle positions on to the image plane.
The Cartesian axes of the plot are aligned with the longest and the shortest principal axes of inertia
of the halo particles and the blue ellipse shows the projection of the inertia ellipsoid. The grey-scale in the background highlights the nested mass resolution levels of the re-simulations and is not relevant to our analysis.
It is worth noticing, however, that the central light-grey area indicates the high-resolution region of the re-simulations which corresponds to the Lagrangian patch occupied by the particles that are found within $3R_{\rm h}$ from the halo centre at $z=0$. For Supay, this volume is very well aligned with the cyan patch while a substantial misalignment is noticeable for Amun. This fact suggests that concentric Lagrangian shells follow a more consistent dynamics in Supay than in Amun.

In the central panels of Fig.~\ref{fig_part_dist}, we plot the particle distribution at $z=0$. The orientation of
the axes is the same as in the left-hand panels but the comoving area shown here is a bit smaller.
The white circumference indicates $R_{\rm h}$ and all the non-halo particles within $\pm 100\,h^{-1}$ kpc in the direction perpendicular to the page have been rendered using grey circles with size proportional to the particle mass and fixed transparency. 
These plots show that Amun forms a node in the cosmic web and is 
connected to a series of thin filaments. Some of these are more prominent than others but there is clearly no dominant structure surrounding the halo.
On the contrary, 
Supay is embedded in a very pronounced filament which is much thicker than $R_{\rm h}$.
Note that the filament is perfectly aligned with the shortest principal axis of the protohalo and perpendicular to the longest one. The entire high-resolution region has been compressed into the filament.
Similar visual examples of the connection between halo assembly and the cosmic web have been published by \citet[][see their Fig. 2]{Hahnetal2009}.
Their statistical study based on large cosmological simulations shows that the halo mass-assembly rate correlates most strongly with the largest eigenvalue of the rate-of-strain tensor (a measure of velocity shear) which reflects the integrated effect of gravitational tides over time.
What is still missing, however, is a detailed study of a few examples that help establishing physical links between these correlated variables: how does matter accretion 
respond to the more or less sheared dark matter flows surrounding haloes?
We address this issue below by analysing our zoom simulations. 

To study how matter accretes to the haloes,  we hand-pick a set of particles based on their Lagrangian
coordinates and follow their trajectories. In particular, we consider the material that is initially contained within
a right circular cylinder symmetrically located around the protohalo centre of mass. The radius of the cylinder
is small compared to its height which is comparable with the size of the protohalo. In Fig.~\ref{fig_part_dist}
we highlight the material in the cylinder with colours indicating its distance from the protohalo centre (left-hand panels) and also show its final Eulerian position at $z=0$ (middle panels).
We finally select 21 individual particles initially located within the cylinder (and equally spaced in $\sqrt{r}$ to put more emphasis on the halo boundary) and plot the time evolution of their distance to the centre of the
halo main progenitor
using the same colour coding (right-hand panels). As a reference, we also show $R_{\rm h}$ as a function of time as black line. 
For Amun (top row), we orient the cylinder with a slight tilt (15$^\circ$ inclination) with respect to the shortest principal axis of inertia of the protohalo. 
Of course, the particles that were initially found close to the centre of the cylinder end up forming the halo. 
Those that originally are in the outskirts, instead, still form a continuous sequence at $z=0$ which is aligned with the short principal axis of inertia of the protohalo. 
This implies that the particles follow nearly radial orbits with respect to the halo centre.
The top-right panel of Fig.~\ref{fig_part_dist} shows that the most distant particles are approaching or have just experienced turn around at the present time. 
They will then accrete to the halo in the future.
In the left-hand panel of Fig.~\ref{fig_halo_acc}, we colour the simulation particles at $z=0$ according to their radial velocity: blue tones indicate infall while red ones correspond to outflows. 
The thin white cloud marks the locus of the particles that are turning around. 
Matter flows towards Amun along a few preferential directions which are isotropically distributed around the halo.
To first approximation, the collapsing patch can be described as an ellipsoid (as in many theoretical models for halo collapse, see Section \ref{sec_model}) 
although the distance of the particles that are turning around is always substantially larger than $2 R_{\rm h}$ (which is what is generally assumed).

We now repeat the analysis for Supay (middle row of Fig.~\ref{fig_part_dist}), once again considering a cylinder which is slightly tilted with respect
to the short principal axis of inertia of the protohalo. In this case, this direction also coincides with the orientation of the thick filament embedding the halo at $z=0$. 
All the particles that are not part of the halo form a thin plume
of matter lying at the centre of the filament at the present time. Most of them, however, are found at very large distances from the halo (i.e. outside the range of $\pm 1.5\, h^{-1}$ Mpc shown in the figure). 
This is a consequence of the fact that the velocity field in the filament has a very strong shear and matter recedes from
the halo as clearly demonstrated in the right-hand panel of Fig.~\ref{fig_halo_acc}.
Further evidence is provided by the radial trajectories of the particles that
show a clear separation between those that are orbiting within Supay and those that are receding from it (middle-right panel in Fig.~\ref{fig_part_dist}). 
We conclude that matter cannot accrete to the halo along the filament.

Since the protohalo of Supay is very asymmetric and elongated, it is interesting to replicate our analysis using
a different set of particles. We thus consider a narrow cylinder which is slightly tilted with respect to the
first principal axis of inertia of the protohalo and perpendicular to the final filament (bottom panels
in Fig.~\ref{fig_part_dist}).
At $z=0$ this material is stretched into a thin strip of matter that runs along the filament. Particles that
initially were lying on the left-hand side of the protohalo are now found on the right-hand side of the filament and vice versa. 
This phenomenon reveals that the material originally contained in the cylinder underwent orbit crossing during the formation of the thick filament which is a fully non-linear structure.
Examining the radial trajectories of the particles, we notice that also in this case there is a clear gap in the final position between the particles bound to the halo and those dispersed along the filament.
Material that immediately surrounds the protohalo in the initial conditions is now receding from the halo (sometimes after having experienced turn around followed by an episode of positive acceleration). 
We also note that some particles found initially at the boundary of the protohalo, nowadays follow orbits that loiter at approximately fixed distance from the halo centre with a dominant velocity component in the tangential direction. They form the outermost shell of matter that can accrete to the halo.

\citet{Hahnetal2009} have shown that the velocity shear of the dark matter flow surrounding a halo correlates with the tides due to the dominant neighbour halo.
These authors used  the restricted three-body problem to quantify the tidal influence of larger haloes.
Consider two bodies of mass $M$ (the primary mass) and $m\ll M$ (the secondary mass) orbiting each other at distance $d$.
The gravitational region of influence of the secondary mass is approximately bounded by the Hill radius, $R_{\rm H}\simeq d\, [m/(3M)]^{1/3}$, i.e. the maximum distance
at which a test particle can stably orbit the secondary without being pulled away by the primary (this is a proxy for the size of the complex Roche lobes).
To extend the applicability of this concept in the presence of multiple neighbours, \citet{Hahnetal2009} considered a given halo as the secondary body and determined the corresponding minimal Hill radius by varying the primary among all more massive objects. 
They found that the mass flux into the minimal Hill sphere is reduced for the least accreting haloes which provides a hint towards tidally suppressed mass growth.
Repeating this analysis, we find that the minimal Hill radii of the stalled haloes are substantially smaller than for the accreting ones
(see Table~\ref{tab_coll_a}) which is not surprising as they cluster much stronger. 
In all cases, however, the minimal Hill radius is much larger than $R_{\rm h}$ and $R_{\rm sp}$ showing that the tidally dominant neighbour is not directly responsible for the stalled accretion (nor it causes any mass loss).
\citet{Hahnetal2009} envisaged that stalled haloes can accrete material lying within their minimal Hill radii as long as it does not move with too large relative velocities. 
In the right-hand panel of Fig.~\ref{fig_halo_acc}, we plot the radial-velocity map for the matter surrounding Supay at $z=0$. 
Infall takes place within a donut shaped region that extends till the turnaround radius (which is comparable with $R_{\rm H}$) while strong and coherent outflows are present within the filamentary structure that hosts Supay.
In this case, the tidally dominant halo has a mass of $5.41\times 10^{12}\,h^{-1}$ M$_\odot$ and lies at a distance of $3.01\,h^{-1}$ Mpc from Supay mostly in the direction of the filament. This halo is actually receding from Supay: both objects are flowing towards a very large mass concentration located at one extreme of the filament but Supay, which
is more distant from it, moves more slowly.
The second and third most dominant haloes are two much more distant cluster-sized objects ($M_{\rm h}\simeq 10^{14}\,h^{-1}$ M$_\odot$, $d\simeq15\,h^{-1}$ Mpc) 
which lie in opposite directions with respect to Supay and, substantially, define the filament.
In general, our high-resolution simulations reveal that reasoning in terms of the minimal Hill sphere provides a poor description of the dark matter flow around stalled haloes mainly
because of the strong geometrical asymmetry due to the presence of elongated large-scale structures.

\begin{figure}
 \includegraphics[width=0.49\textwidth,bb=0 0 1074 700,keepaspectratio=true]{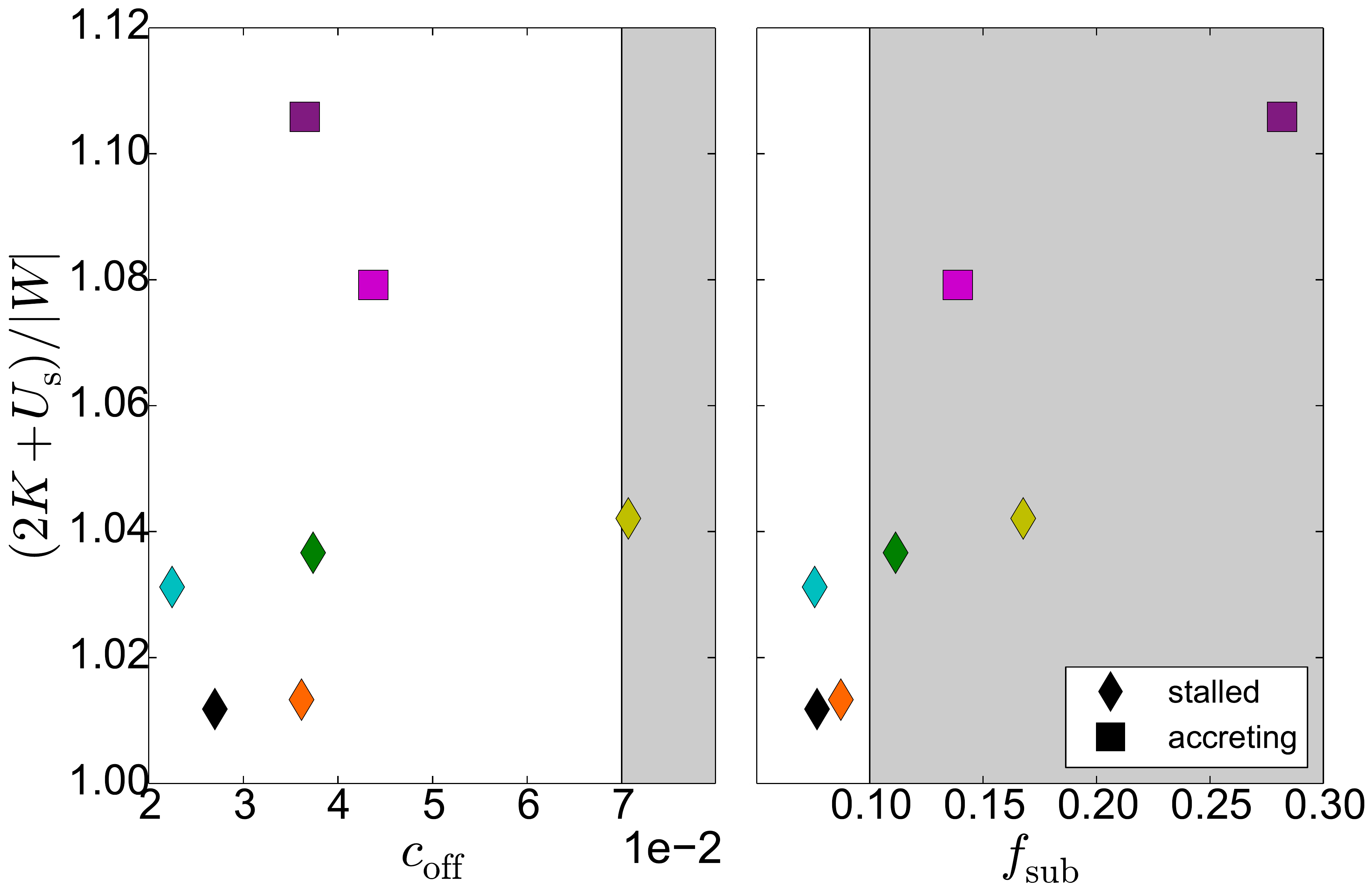}
 \caption{Indicators of the dynamical state of the resimulated haloes at $z=0$:
the virial ratio $(2K+U_{\rm s})/|W|$ (see the main text for details), the offset between the centre of mass and the location of the potential minimum in units of $R_{\rm h}$, $c_{\rm off}$, and the mass fraction in resolved substructures, $f_{\rm sub}$.
 The unshaded area indicates the region of 
 parameter space occupied by dynamically relaxed halos 
 ($(2K+U_{\rm s})/|W|<1.35$, $c_{\rm off}<0.07$, $f_{\rm sub}<0.1$) according to the study by
  \citet{Netoetal2007} which is based on large cosmological simulations with a lower
  mass resolution than ours and thus provides smaller values for $f_{\rm sub}$.
 Overall, stalled haloes appear to be more relaxed (i.e. closer to dynamical equilibrium) with respect to the accreting ones.}
 \label{fig_relax}
\end{figure}

\begin{figure*}
\includegraphics[width=0.49\textwidth,bb=0 0 688 607,keepaspectratio=true]{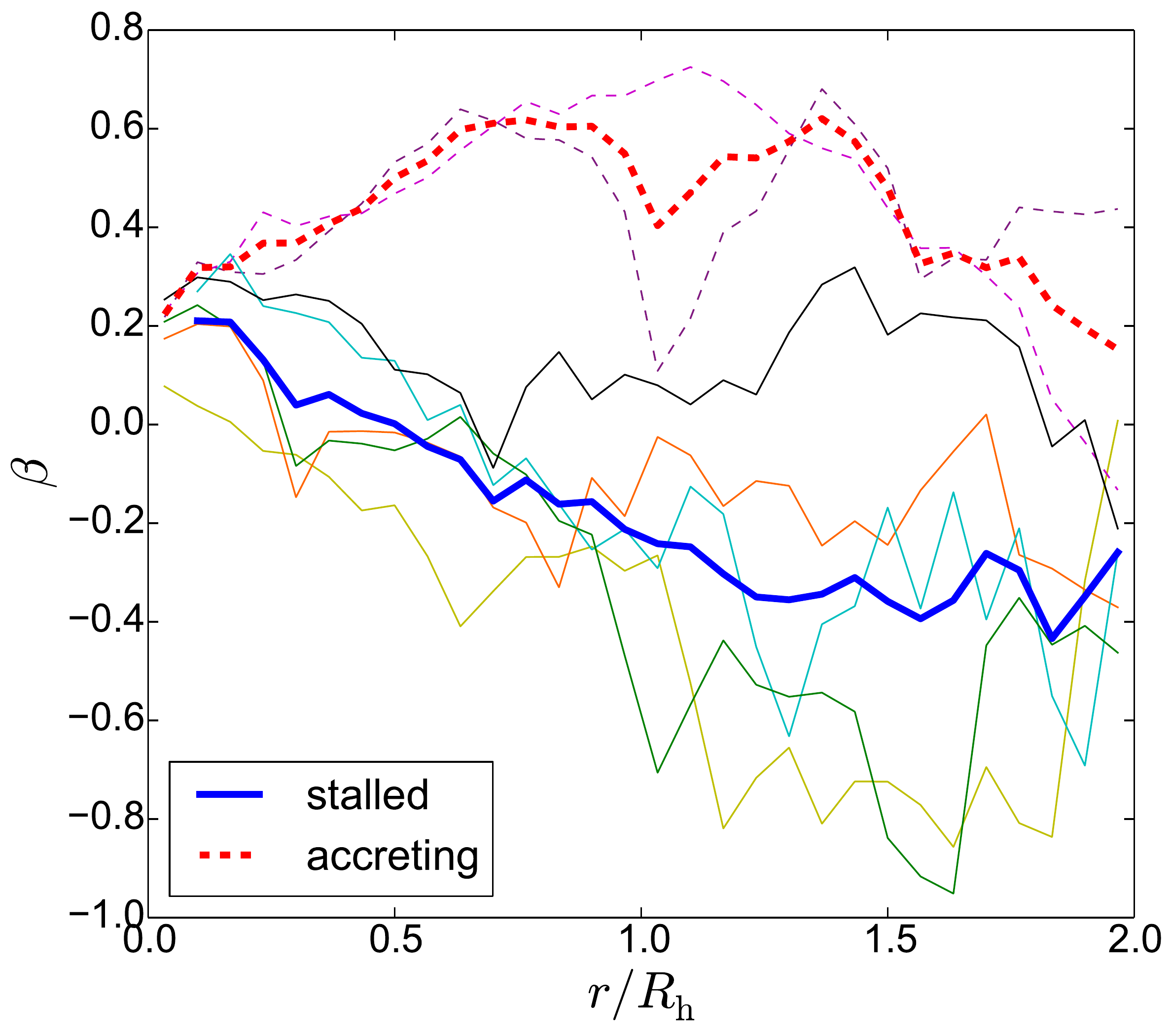}
\includegraphics[width=0.49\textwidth,bb=0 0 687 607,keepaspectratio=true]{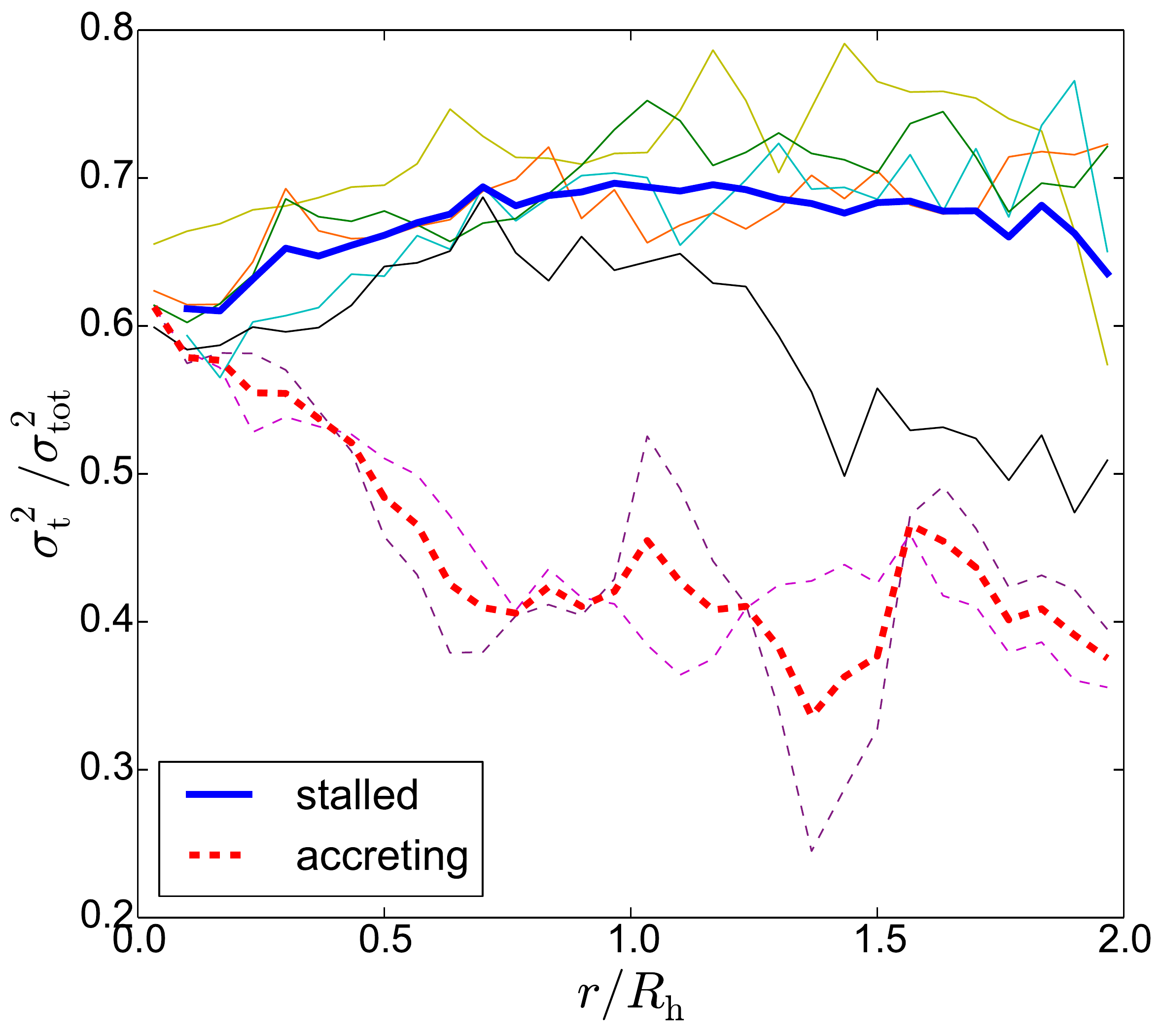}
\caption{The present-day velocity anisotropy parameter $\beta$ (left-hand panel) and the ratio of tangential to total velocity variance (right-hand panel) measured in thin spherical shells at distance $r$ from the centre of our resimulated haloes. 
The thick solid and dashed lines show averages taken over all the stalled and accreting haloes, respectively.
The thin lines refer to the individual haloes (with the usual colour coding).
Circular orbits play a much more important role in the stalled haloes due to the impact of the host filament on halo assembly.}
\label{fig_velocit_anisortopy}
\end{figure*}

\subsection{The impact of the cosmic web on halo structure}\label{sec_struct} 
We now focus on the material that forms the haloes at the present time and explore differences in the
internal structure of accreting and stalled haloes.

We first consider halo shapes at both the initial and final times.
As evidenced in Table~\ref{tab_coll_a}, Fig.~\ref{fig_part_dist}, and briefly described at the beginning of Section~\ref{sec_results}, 
the Lagrangian patches of stalled haloes show a characteristic elongated shape in which
the shortest principal axis of inertia aligns with the filament (see also \citealt*{Porcianietal2002}; \citetalias{Ludlow2014}). 
This phenomenon reflects the fact that halo growth is inhibited along the filament  by a strong
velocity shear while material can more easily flow in along the perpendicular directions (this is the way the filament itself is formed, actually).
As already mentioned, at $z=0$, no obvious difference can be noticed in the halo shapes between stalled and accreting haloes (see Table~\ref{tab_coll_a}). 
However, the minor axis of stalled haloes retains a `memory' of the halo-assembly process and is preferentially oriented in the perpendicular direction to the filament.
The mass-density profiles of the resimulated haloes at $z=0$ are well approximated by the NFW formula. 
With the exception of Seth, stalled haloes tend to be slightly more concentrated.  

Three quantities are often used to infer whether a halo is dynamically relaxed or not:  
the offset between the centre of mass and the minimum of the gravitational potential 
measured in units of the halo radius, $c_{\rm off}$, 
the fraction of mass in resolved substructures, $f_{\rm sub}$, and the virial ratio, $(2K+U_{\rm s})/|W|$.
(Here, $K$ and $W$ denote the kinetic and the potential energy of the halo, respectively, while $U_{\rm s}$ is the correction due to the surface pressure of the particles crossing the halo boundary.)
In Fig.~\ref{fig_relax}, we contrast the accreting haloes (squares) with the stalled ones (diamonds) in terms of
these quantities.
Accreting haloes show a systematically higher virial ratio and a tendency towards having larger $f_{\rm sub}$  
while no significant difference in $c_{\rm off}$ is seen. Seth distinguishes itself from the other stalled haloes
by showing the largest values of the virial ratio, $f_{\rm sub}$ and, in particular, $c_{\rm off}$.
This is because it contains a massive substructure (with mass ratio 1:17) that just had its first pericentre passage. 

The internal dynamics of the haloes can be conveniently described in terms of the velocity anisotropy parameter,
\begin{eqnarray}
 \beta=1-\frac{\sigma_{\rm t}^2}{2\sigma_r^2},
\end{eqnarray}
where $\sigma_r$ and $\sigma_{\rm t}=(\sigma^2_\theta+\sigma^2_\phi)^{1/2}$ denote the radial and tangential velocity dispersions, respectively.
An isotropic velocity distribution has $\beta=0$, while $\beta<0$ indicates a bias towards tangential motions. 
\citet{Faltenbacher2010} showed that classifying haloes in terms of $\beta$ leads to a very strong assembly bias. Stacking thousands of haloes from a large cosmological simulation, 
they also found a systematic difference between the radial $\beta$ profiles of halo subpopulations split based on some other halo properties (e.g. shape or spin) at fixed halo mass.
These unexplained results suggest that the halo velocity structure correlates with the large-scale environment and
provide us with the motivation to analyse what happens in single objects using our zoom runs.
In the left-hand panel of Fig.~\ref{fig_velocit_anisortopy}, we report the individual and averaged radial $\beta$ profiles 
extracted from our re-simulations. Stalled and accreting haloes show consistent dynamics only in their innermost cores (i.e. for $r/R_{\rm h}<0.15$ where $\beta\simeq 0.2$).  
Further out, their orbital structure is very different. Within $R_{\rm h}$, $\beta$ increases with $r$
for the accreting haloes while it decreases and even turns negative for the stalled ones that are
therefore dominated by tangential motions in their outer regions.
The origin of this phenomenon can be schematically understood as follows. 
To first approximation, matter falls towards accreting haloes
following radial orbits as it is assumed in the spherical or ellipsoidal collapse models.
However, this is not the case for the stalled haloes.
In fact, the presence of the host filament has a dramatic influence on the orbits of the surrounding material particles
that get deflected from the radial direction and accrete to the halo with a significant tangential-velocity component.
Due to the approximate symmetry around the filament axis, both senses of rotation are possible for the accreting particles and stalled haloes do not spin-up substantially
(see Table~\ref{tab_coll_a}). Small asymmetries in the accretion flow, however, could potentially generate a spin component parallel to the filament.
For our six re-simulated stalled haloes, this component is always subdominant (the mean and standard deviation of the angle between the spin and the filament direction
are $66^\circ$ and $17^\circ$, respectively).
The velocity anisotropy also reflects changes in the radial-velocity dispersion which 
is slightly higher for the accreting haloes due to the contribution of their radially infalling streams.
In order to get a more complete picture, we
show the tangential contribution to the total velocity variance in the right-hand panel of Fig.~\ref{fig_velocit_anisortopy}. 
Once again the differences between stalled and accreting haloes are striking and consistent with our previous findings. 
We conclude that  the dynamical effects due to the gravitational attraction of the filament are already noticeable deep inside the haloes where stalled haloes are dominated by tangential orbits and accreting haloes by radial ones.
This intimate link between the velocity anisotropy parameter and the formation history of the haloes
elucidates the results by  \citet{Faltenbacher2010}.
Our results complement (and are consistent with) several recent studies on the complex interplay between the cosmic-web dynamics and the spins of galaxy-sized haloes
(independently of their assembly histories).
Orbit-crossing generates vortical flows within multistream regions  \citep{Pichon-Bernardeau-1999, Pueblas-Scoccimarro-2009, Libeskindetal2014, Wangetal2014} and 
the resulting vorticity is preferentially perpendicular to the axis along which gravitational collapse proceeds faster \citep{Libeskindetal2013, Laigleetal2015}.
In the plane perpendicular to a filament, vorticity shows a characteristic quadrupolar pattern \citep{Laigleetal2015} originated by the winding of the matter flows \citep{Pichonetal2011, Codisetal2012}. Halo spins are thus influenced by secondary anisotropic infall of matter from the vortical filaments \citep{Aubert-Pichon-Colombi2004, Bailin-Steinmetz2005, Codisetal2012,Libeskindetal2013}.
As a consequence, the angular momentum of low-mass haloes ($M_{\rm h}<5\times 10^{12}$ M$_{\odot}$ at $z=0$) tends to
be preferentially aligned (in a statistical sense) with the cosmic-web filaments in which they reside, whereas the spin of larger objects lies in the perpendicular direction
\citep{Hahnetal2007a, Aragon-Calvoetal2007, Sousbieetal2008, Pazetal2008, Zhangetal2009, Codisetal2012, Aragon-Calvo-Yang2014}.
\citet{Laigleetal2015} make the conjecture that these correlations reflect the Lagrangian size of the haloes compared with the Lagrangian stretch of the vorticity quadrants forming within
the filament. Small haloes feel only one polarity and get an important spin component along the filament.
On the other hand, more extended objects overlap with quadrants of opposite polarity and do not develop a net angular-momentum component parallel to the filament, $L_\parallel$.
In order to clarify what happens in our simulations, we compute the specific $L_\parallel$ per unit mass for the particles immediately surrounding the stalled haloes (i.e. lying between $R_{\rm h}$ and $R_{\rm spl}$).
The spatial distribution of the specific $L_\parallel$ in the plane perpendicular to the filament presents a complicated pattern (approximately quadrupolar with superimposed small-scale fluctuations), while its azimuthally averaged probability density is broad and approximately symmetric for the two senses of rotation.
The mean value corresponds to 20-30 per cent of the mean specific angular momentum of the halo particles. 
The same result is found considering only infalling material.
This suggests that, in the mass range we consider, accretion from the filament should not alter the halo spin significantly.
The kinematic structure of the filament mostly imprints a tangential-velocity dispersion in the halo as a characteristic signature.

\begin{figure}
 \includegraphics[width=0.5\textwidth,bb=0 0 716 661,keepaspectratio=true]{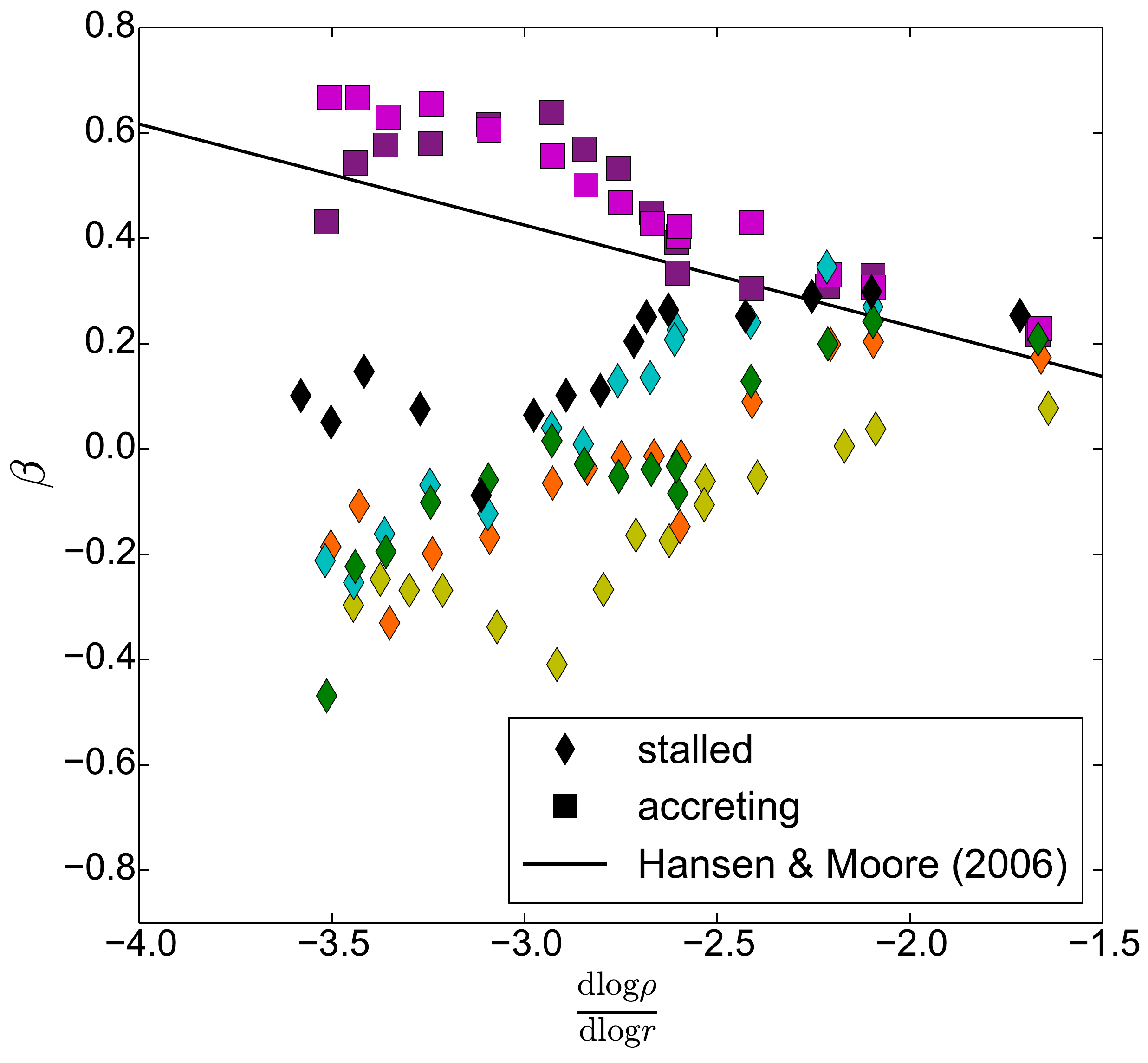}
\caption{The velocity anisotropy parameter measured in thin radial shells for our resimulated haloes at $z=0$
is plotted against the corresponding slope of the mass-density profile.
 Only shells contained within $R_{\rm h}$ are considered.
 The solid line indicates the universal relation proposed by \protect\cite{HansenMoore2006}.}
\label{fig_beta_gamma}
\end{figure}

\begin{figure*}
 \includegraphics[width=0.95\textwidth,bb=0 0 1152 576,keepaspectratio=true]{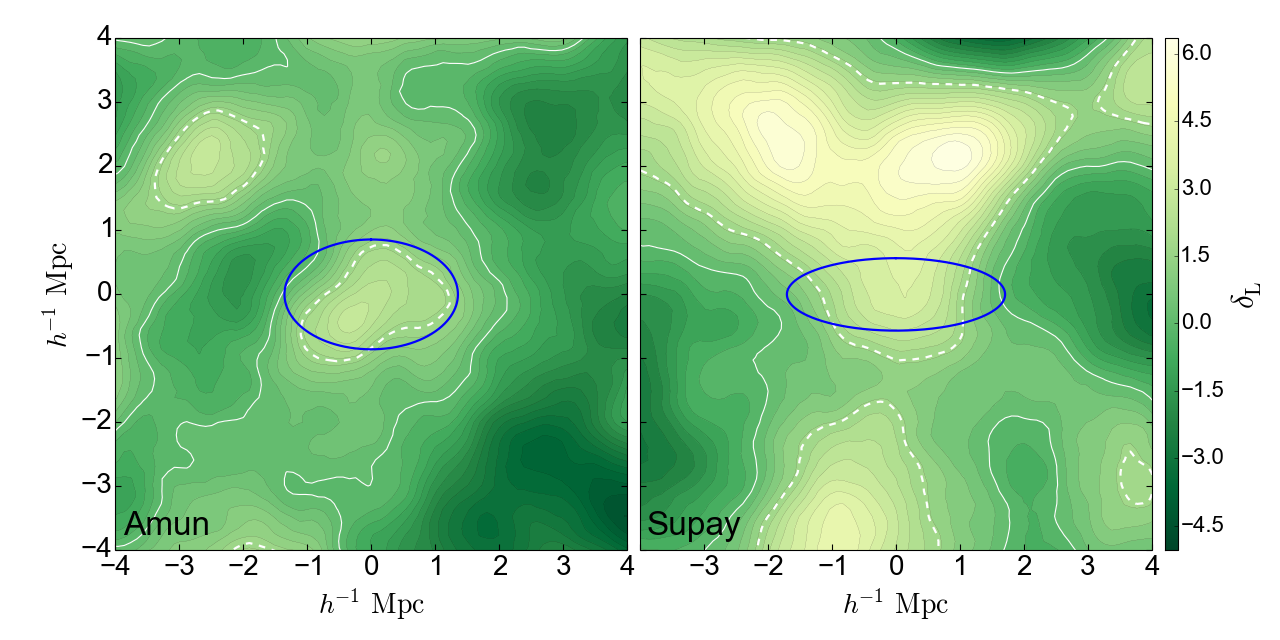}
 \caption{Map of  the linear density field at $z=0$ on a slice passing through the protohalo centres of Amun (left) and Supay (right).
 The orientation of the axes is as in Fig.~\ref{fig_part_dist}.
 The solid white contour lines indicates the mean density of the universe while the dashed one corresponds to the spherical-collapse threshold $\delta_{\rm L}=1.686$.
 The blue ellipse is the projection of the inertia ellipsoid of the protohalo as in Fig.~\ref{fig_part_dist}.
 The density contrast is smoothed with a top-hat filter in configuration space with mass resolution $M_{\rm s}=M_{\rm h}$.}
 \label{fig_linear_density}
\end{figure*}

Early studies suggested that $\beta$ follows a universal radial profile in relaxed haloes \citep*[e.g.][]{ColeLacey1996,Carlberg1997,Rasiaetal2004}.
More recently, \citet{HansenMoore2006} provided evidence for
a universal relation between $\beta$ and the slope of the radial mass-density profile (both evaluated
at the same position).
Our simulations are inconsistent with such a relation. 
In fact, while the density profiles of all haloes are monotonically decreasing, the $\beta$ profiles 
increase with $r$ for accreting haloes and decrease with $r$ for stalled haloes.
The velocity anisotropy and the logarithmic slope of the density profile for different radial shells 
(all with $r<R_{\rm h}$) of
our re-simulated haloes are shown in Fig.~\ref{fig_beta_gamma}. 
The solid line highlights the fit derived by \citet{HansenMoore2006}.
Accreting haloes approximately follow the assumedly universal relation while stalled haloes violate it. 
In both cases, better agreement with the fit is found
for the innermost shells which are displayed on the right-hand side of the figure.
We deduce that the universal relation proposed by \citet{HansenMoore2006} holds true only in the 
halo core ($r<0.2 R_{\rm h}$) where no difference is found between accreting and stalled haloes.
Several authors have reached similar conclusions following different approaches
 \citep[e.g.][]{Navarroetal2010,Ludlowetal2011,Lemzeetat2014} even in the presence of
 a baryonic-matter component \citep{Tisseraetal2010}.
The reason why \citet{HansenMoore2006} could not notice any departure from a universal relation
is that they considered a special set of simulations: 
single overdensities undergoing spherical collapse, two colliding objects, and a very massive halo extracted from a cosmological simulation. All these systems show a continuous radial infall of matter at late times and would be tagged as accreting in our classification scheme.

\begin{figure}
 \centering
 \includegraphics[width=0.47\textwidth,bb=0 0 737 737,keepaspectratio=true]{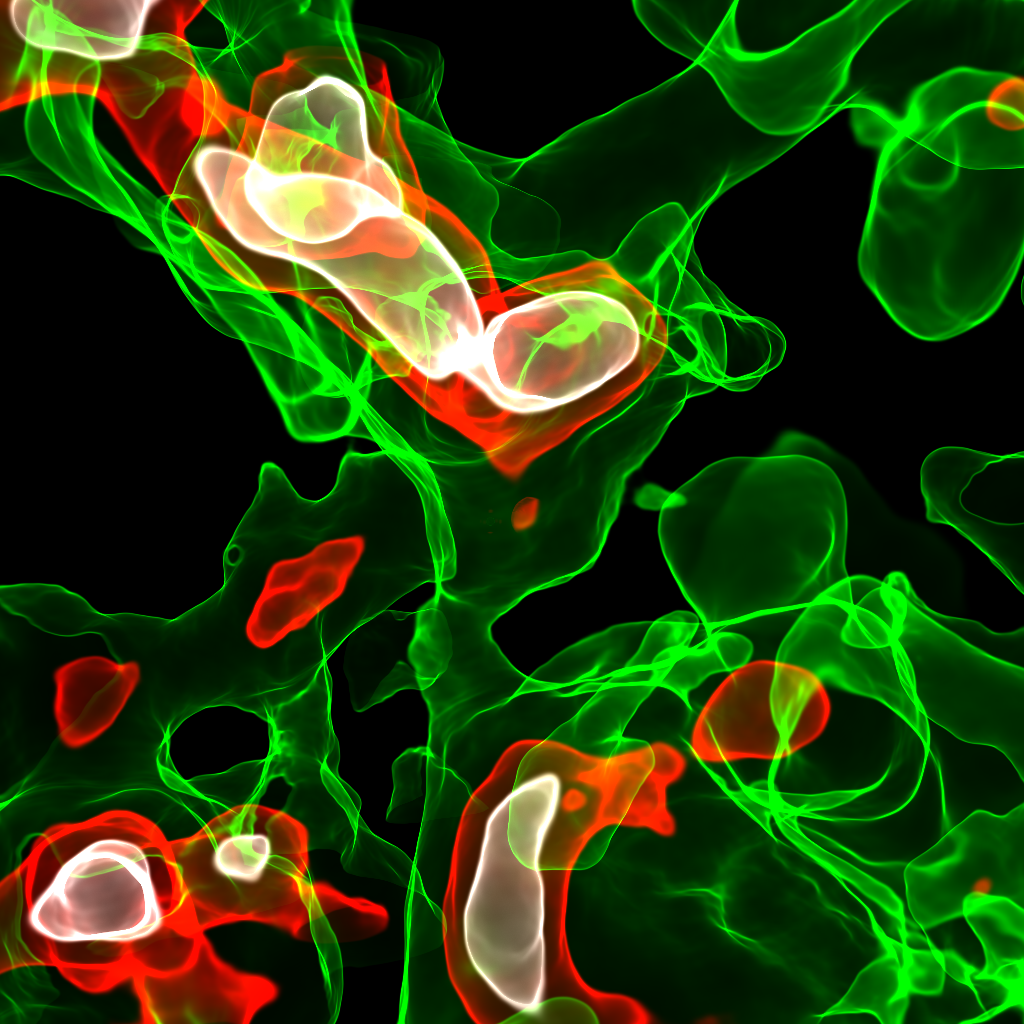}
 \caption{Volume rendered image of the linear density field presented in the right-hand panel of Fig.~\ref{fig_linear_density}. 
 Selected isodensity surfaces with $\delta_{\rm L}=1.68, 3.9$ and 4.9 are drawn in green, red and white, respectively. 
 Supay originates from the small red peak located at the centre of the figure.}
 \label{Supay-rendering}
\end{figure}

\subsection{Predicting halo masses with the excursion-set method}\label{sec_model}
Analytical methods that establish a link between dark matter haloes and the properties of their Lagrangian
patches in the linear density field are extremely useful in cosmology as they provide the tools to make predictions for the halo mass function and clustering properties.
In principle, the rarity and biasing of density peaks in different environments of the cosmic web follows directly from the statistics of Gaussian random fields \citep{Doroshkevich1970}.
However, for any practical application, it is necessary to express the results as a function of the final halo properties (rather than the initial peak height, etc.).
The challenge is to create a mapping from the initial to the final configurations in terms of a collapse model (e.g. \citealt{PeeblesBook1980,BM1996}; \citetalias{Ludlow2014}) and deal with the multi-scale nature of the density field.
The most successful algorithm is certainly the excursion-set method \citep{BCEK, Zentner}.
In this section we test it against our high-resolution simulations.

We extract the linear density field from the initial conditions of the parent U512 simulations
and use the linear growth factor of matter perturbations to evaluate the linear density contrast
at the present time.
We convolve this field with a top-hat kernel in configuration space and obtain the smoothed linear density
contrast, $\delta_{\rm L}$. We vary the smoothing radius, $R_{\rm s}$ and associate it with the mass scale $M_{\rm s}=\frac{4\pi}{3}R_{\rm s}^3\overline{\rho}_{\rm m}$ where $\overline{\rho}_{\rm m}$ denotes the mean matter density in the universe.
For example, in Fig.~\ref{fig_linear_density} we draw maps of $\delta_{\rm L}$ for Amun and Supay with resolution $M_{\rm h}$ (we use a plane
passing through the protohalo centre of mass with the same orientation as in Fig.~\ref{fig_part_dist}). 
To facilitate understanding, in Fig.~\ref{Supay-rendering} we also present a volume rendering of the linear density field for Supay.
Both haloes originate from a density peak (although, for Supay, this is not noticeable in Fig.~\ref{fig_linear_density} as the local maximum is slightly off the plane of the image).
However, while Amun emerges from a rather isolated peak, the protohalo of Supay lies in between two extended and very dense structures
(its large-scale layout closely reminds that of a saddle point with a superimposed small-scale peak).
Only such particular configurations in the linear density field lead to the formation of strong filaments.
What is necessary is the presence of two rare high-density peaks that are not too distant from each other and whose tidal tensors are sufficiently well aligned \citep{BKP}. 
This layout evolves into two massive haloes joint by a thick filament.
Consistently with this concept, our stalled haloes always lie within 2--3$\,h^{-1}$Mpc from a more massive object and move along the filament. 
The relative velocities indicate that merging might only happen in some distant future thus confirming that stalled haloes are long lived objects.

\begin{figure}
 \includegraphics[width=0.49\textwidth,bb=0 0 729 585,keepaspectratio=true]{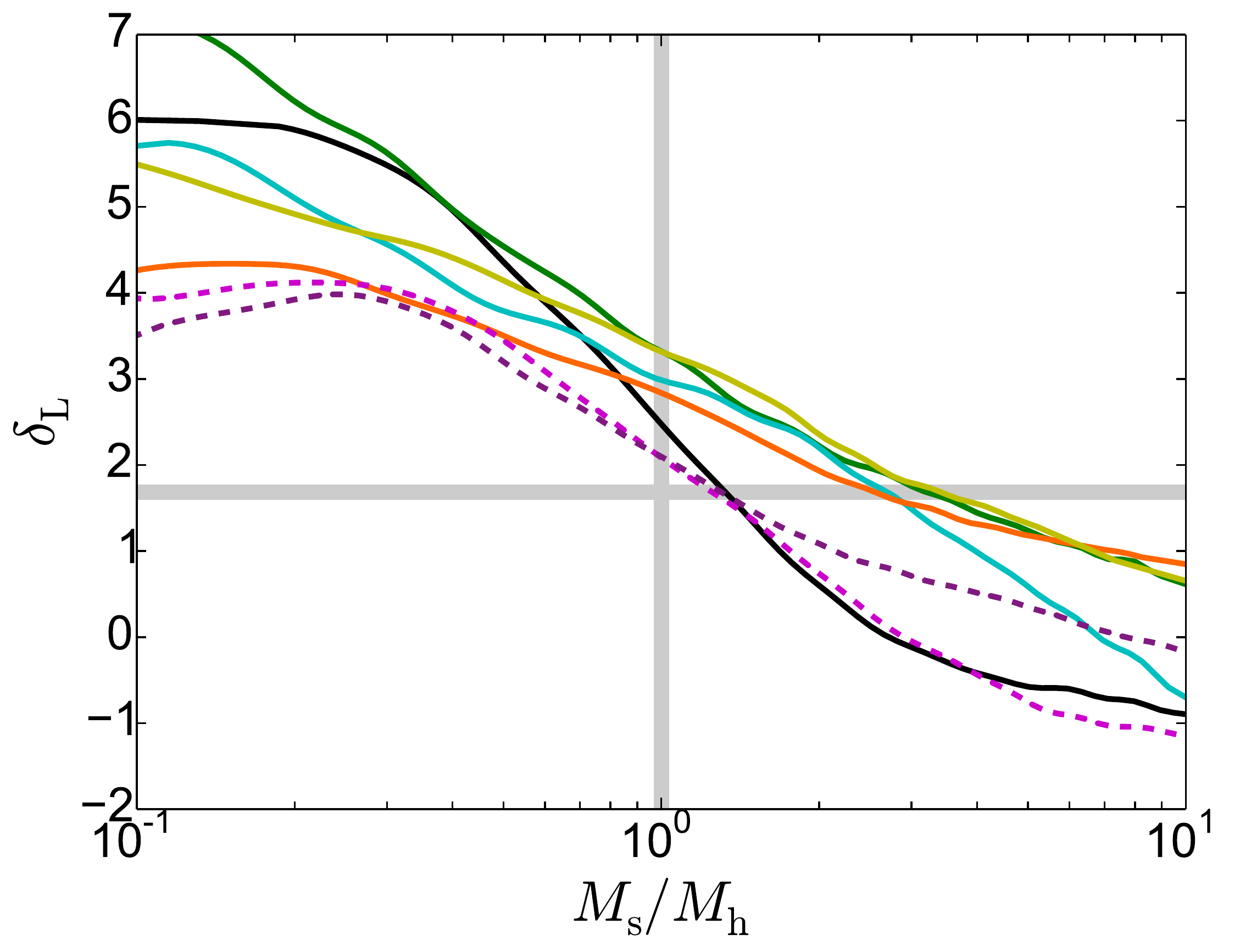}
\caption{The linear density contrast evaluated at the centre of the D-ZOMG protohaloes 
is plotted as a function of the smoothing mass scale for a spherical top-hat filter in configuration space.
The vertical and horizontal 
grey lines mark the mass of the haloes in the $N$-body simulations and the spherical-collapse threshold,
respectively.
Note that collapse time correlates very well with the linear density contrast measured on the halo mass scale, $M_{\rm h}$.}
\label{fig_linear_dens}
\end{figure}
\begin{figure}
 \includegraphics[width=0.49\textwidth,bb=0 0 729 585,keepaspectratio=true]{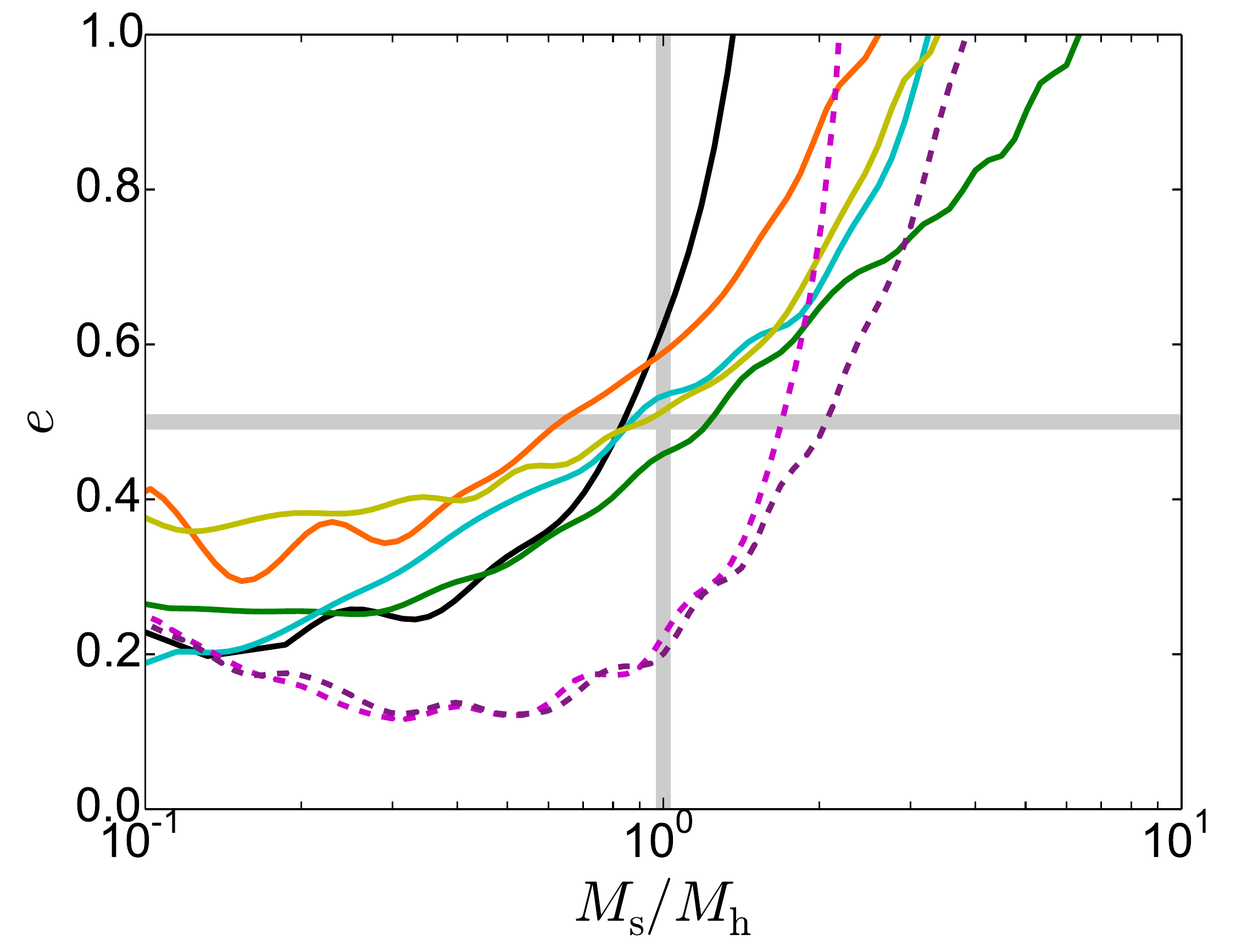}
\caption{As in Fig.~\ref{fig_linear_dens} but for the ellipticity of the linear tidal field instead of the density contrast.
The vertical and horizontal grey lines indicate the mass of the haloes in the $N$-body simulations and
the value $e=0.5$, respectively.
Note that the environment of stalled haloes produces stronger tides.}
\label{fig_linear_ellip}
\end{figure}

Excursion-set trajectories are obtained by plotting $\delta_{\rm L}$ evaluated at the Lagrangian positions of the haloes versus $M_{\rm s}$.
In Fig.~\ref{fig_linear_dens}, we show the trajectories for the D-ZOMG haloes and compare them with
the collapse threshold derived from the spherical collapse model (horizontal grey line).
We first note that the slopes of the excursion-set trajectories are pretty similar for accreting and stalled haloes (with the only exception of Siris) thus indicating that their protohaloes have similar radial density profiles.   
We attempt to predict the final halo masses by identifying the mass scale at which the trajectories
first up-cross the threshold level $B_{\rm SC}=1.686$. This corresponds to requiring that
the outermost mass shell collapses at the present time according to the spherical collapse model \citep{PeeblesBook1980}.
Such a guess overestimates $M_{\rm h}$ by a factor of $\sim 3$ for the stalled haloes while it gives better results (offset by nearly 25 per cent) for the accreting ones.
In addition, Fig.~\ref{fig_linear_dens} shows a tight correlation between the linear density contrast smoothed on the halo mass scale (highlighted by a vertical grey line) and the collapse time as already evidenced by \citetalias{Ludlow2014} and \citetalias{Borzyszkowski2014} using large cosmological simulations. Therefore it is not possible to accurately predict the halo masses using the same threshold for all the trajectories.

\begin{figure*}
 \includegraphics[width=0.95\textwidth,bb=0 0 1448 539,keepaspectratio=true]{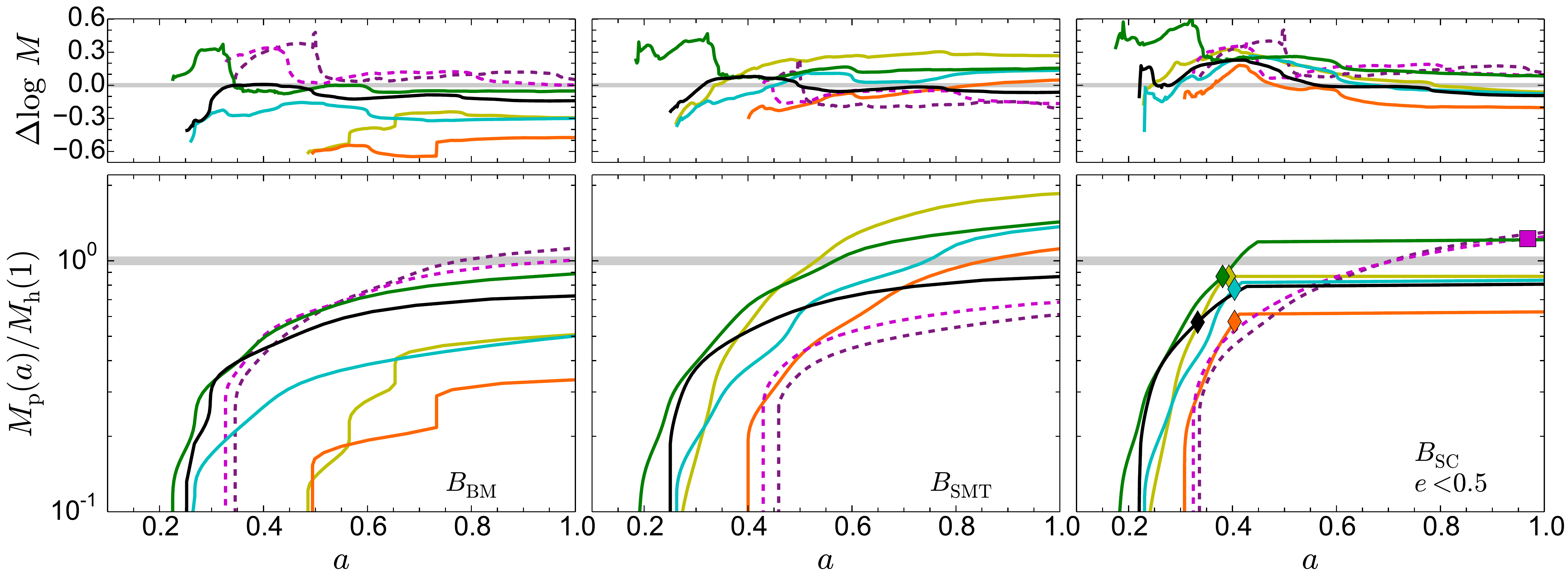}
\caption{Mass accretion histories for the D-ZOMG haloes predicted using the excursion-set method
with different collapse thresholds normalized to the individual measured halo present mass.
The left-hand panel shows results derived with the ellipsoidal collapse model by {\protect \citet{BM1996}}.
The middle panel is based on the fit by {\protect \citet{SMT2001}}.
Finally, in the right-hand panel we combine the spherical-collapse threshold with the additional condition $e<0.5$ for a shell to collapse.
The horizontal band highlights the halo mass measured in the Z4096 simulations at $z=0$ while the
symbols in the right-hand panel indicate the collapse time of the haloes estimated from the same runs. 
The smaller panels on the top show the logarithm of the ratio between the different excursion-set predictions and the mass accretion histories measured in the simulations (see Fig.~\ref{fig_mah_resim}).}
\label{fig_mass_prediction}
\end{figure*}

We now consider additional variables to account for the effect of tides that we quantify in terms of the linear deformation tensor, $D_{ij}=\partial_i\partial_j \phi$,
where $\phi$ is the gravitational potential obtained solving $\nabla^2 \phi=\delta_{\rm L}$.
After computing the eigenvalues of the deformation tensor, $\lambda_1\geq \lambda_2 \geq \lambda_3$, we define the ellipticity of the tidal field as
\begin{eqnarray}
 e=\frac{\lambda_1-\lambda_3}{2\delta_{\rm L}}
\end{eqnarray}
(we recall that  $\lambda_1+\lambda_2+\lambda_3=\delta_{\rm L}$).
This quantity measures the degree of anisotropy between the most compressing and stretching components of the linear tidal field (or of the linear velocity shear).
In Fig.~\ref{fig_linear_ellip}, we show how $e$ (evaluated at our protohalo centres) changes as a function of the smoothing scale.
For $M_{\rm s}\simeq M_{\rm h}$, stalled haloes experience significantly stronger tides than accreting ones.
Information on the formation of a filament in their surroundings is therefore already encoded in the linear
density field.

The ellipsoidal model by \citet{BM1996} is commonly used to describe gravitational collapse in the presence
of tides. In this case, an initially spherical patch is distorted into an ellipsoid by the action of linear tides and
collapse takes place at different times along the different principal axes. As a matter of fact, collapse along
all the three axes corresponds to an excursion-set threshold, $B_{\rm BM}$, that depends on $\lambda_1, \lambda_2$ and $\lambda_3$. 
We apply this model to our protohaloes and we show the resulting mass-accretion histories in the left panel of Fig.~\ref{fig_mass_prediction} 
(sharp variations are due to the oscillations in $e$ seen in Fig.~\ref{fig_linear_ellip}). 
Predictions for the halo mass at $z=0$ are quite accurate for the accreting haloes but severely underestimate $M_{\rm h}$ for the stalled haloes.
The latter class of objects, in fact, shows very elongated protohaloes that cannot be well approximated with spheres.
Note that the ratio between the predicted and measured mass accretion histories is nearly constant (top-left panel).

\citet*{SMT2001} combined the ellipsoidal collapse model with the statistics of Gaussian random fields 
and derived an effective threshold for halo formation which is a function of $M_{\rm s}$ only. 
The underlying idea is that lower-mass objects experience stronger tides (on average) and thus need higher
initial densities to overcome the tidal forces and collapse by a given time.
Once obtained a functional form for the threshold, \citet{SMT2001} arbitrarily rescaled some of its coefficients
in order to fit the halo mass function and linear bias parameters extracted from $N$-body simulations,
the justification being the freedom in the definition of the halo boundary.
The resulting threshold, $B_{\rm SMT}$, accurately describes the mean protohalo densities found in simulations as a function of halo mass (\citealt{Robertsonetal2009}; \citealt*{Elia}) although, contrary to expectations, the scatter around the mean correlates with the halo collapse or formation time \citepalias{Ludlow2014, Borzyszkowski2014} and does not depend on the tidal prolateness  \citep{HahnParanjape14}.
In the middle panel of Fig.~\ref{fig_mass_prediction}, we show the mass accretion histories deriving from
the excursion-set method combined with $B_{\rm SMT}$. 
On average the model predicts the correct halo masses at $z=0$ but the individual estimates are not accurate. 
This method tends to underestimate the mass of accreting haloes and to overestimate that of stalled haloes. 
Furthermore, stalled haloes grow at a slightly faster pace in the model compared to the simulations (top-centre panel).

$N$-body simulations show that protohaloes are strongly aspherical and their inertia tensors are closely aligned with the tidal shear (\citealt{LeePen, Porcianietal2002}; \citealt*{Despali2013}; \citetalias{Ludlow2014}). 
Supay provides a clear example of this phenomenon (see Figs.~\ref{fig_part_dist} and~\ref{fig_linear_density}).
A modified ellipsoidal collapse model that follows the evolution of ellipsoidal Lagrangian
regions aligned with the linear velocity shear has been introduced in \citetalias{Ludlow2014}
and further discussed in \citetalias{Borzyszkowski2014}. 
This model provides a better match to simulations. 
Also in this case, however, the linear density required for perturbations to collapse by the present time 
underestimates the values measured for protohaloes in $N$-body experiments.
Dropping the assumption that haloes are forming today and imposing that their outermost shell collapsed
at $t_{\rm c}$ nicely reconciles model predictions and simulations and provides a consistent picture \citepalias{Borzyszkowski2014}.
This works well even when the constant threshold of the spherical collapse model is adopted in the
excursion-set calculations.
The key idea is that haloes grow (in approximate agreement with the spherical collapse model) until
the gravitational effect of surrounding structures inhibits further growth.
The problem is that one needs to estimate $t_{\rm c}$ based on the linear field in order to make the model predictive. 
Useful indications can be extracted from Fig.~\ref{fig_linear_ellip}:
stalled haloes are associated with values of $e \sim 0.5$ for $M_{\rm s}\simeq M_{\rm h}$ 
while accreting haloes present much smaller values of $e$. 
This suggests the following condition for halo growth:
\begin{equation}
e<0.5 \ \ \ \Rightarrow \ \ \ \delta_{\rm L}>\lambda_1-\lambda_3\;.
\label{conditioneq}
\end{equation} 
This equation requires that the spherical part of the linear tidal tensor dominates over the anisotropic part.
Note that this condition is independent of $\lambda_1$. In fact, substituting $\delta_{\rm L}$ with
the sum of the eigenvalues of the deformation tensor, it can be re-written as $\lambda_2>-2\,\lambda_3$.
It basically states that, in the presence of one-dimensional dilating deformations (i.e. when $\lambda_2>0$ and $\lambda_3<0$), $|\lambda_3|$ cannot be too large for accretion to take place. Moreover, halo growth is always inhibited if $\lambda_2$ is negative (i.e. when dilation happens along two or more principal axes). 
The latter case should be tested against numerical simulations as all our stalled haloes have $\lambda_2>0$ for $M_{\rm s}=M_{\rm h}$ (as expected in the presence of a filament). It might be possible that an additional condition on the tidal prolateness is necessary to describe the most general configuration. 

To develop an efficient algorithm that correctly identifies the halo masses given the initial conditions, 
we implement equation~(\ref{conditioneq}) within the excursion-set model as follows:
(i) we build two trajectories for each halo: $\delta_{\rm L}$ versus $M_{\rm s}$ and $e$ versus $M_{\rm s}$;
(ii) we look for the first up-crossing of the spherical-collapse threshold (linearly rescaled with the growth factor
for matter perturbations at each time) and label the corresponding smoothing mass with $M_{\rm up}$;
(iii) if $e<0.5$ on the scale $M_{\rm up}$ we say that the halo mass is $M_{\rm up}$, otherwise we keep
the halo mass unchanged compared with the previous time step.
(Note that if the $e$ trajectory crosses the value 0.5 several times, the halo will experience a series of
accreting and stalled phases.)
The resulting mass-accretion histories are shown in the right panel of Fig.~\ref{fig_mass_prediction}.
Both the collapse time and the final masses of the stalled haloes are nicely reproduced by this very simple scheme
(within 11 per cent for $a_{\rm c}$ and 19 per cent for $M_{\rm h}$).
The ratio between the predicted and measured masses is approximately constant over a large period of time (top-right panel).
Stalled haloes, however, show a slight decrease of the mass ratio, which can be attributed to pseudo-evolution in the simulation.

\section{Summary and conclusions}\label{sec_conc}
We have investigated the origin of halo assembly bias using a suite of $N$-body simulations including
a cosmological box and seven zoom runs.
Our analysis focuses on haloes with a mass of a few$\times 10^{11} h^{-1} {\rm M}_\odot$. 
These are the haloes in which star formation is most efficient and that host $L_*$ galaxies at the present time.

Target haloes for re-simulation have been selected based on the collapse time, $t_{\rm c}$, 
introduced by \citetalias{Borzyszkowski2014}.  This quantity measures the time at which the physical volume
occupied by the forming haloes becomes stable and virialization can be established.
Partitioning haloes of a fixed mass based on $t_{\rm c}$ generates a strong assembly bias.
We sampled the tails of the distribution of $t_{\rm c}$ by picking (for re-simulation)
five isolated haloes that form very early on (at redshift $z>1$) and two that have not yet collapsed by $z=0$ in the parent run.
This set forms the D-ZOMG simulation suite.
We have checked that our sample is mostly representative of the full halo population (less than 5 per cent of the haloes in the parent run do not satisfy the isolation criterion).

We studied the formation process and the internal structure of the resimulated haloes as well as the properties of the Lagrangian patches from which they originate.  Our conclusions are as follows.

\begin{enumerate}
\item Matter steadily accretes to late-collapsing haloes which grow in mass. These `accreting' haloes coincide
with nodes of the cosmic web and are fed by the radial infall of matter along thin filaments. 
Substructures and their associated tidal streams orbit the haloes following eccentric trajectories that extend
up to very large distances. 

\item The radial mass-density profile of early-collapsing haloes does not evolve since $t_{\rm c}$. 
These `stalled' haloes are embedded in prominent filaments of the large-scale structure which are thicker than the halo radius.
Matter recedes from the halo along the filament and infall is only possible from the perpendicular directions (see the right-hand panel in Fig.~\ref{fig_halo_acc} for an illustrative example). 
Although some matter always accretes to the halo, the inflow is balanced by outflows along the direction of the filament so that there is no net mass growth.
Stalled haloes form the bulk of the halo population with a few$\times 10^{11} h^{-1} {\rm M}_\odot$ at $z=0$ (see Fig.~\ref{fig_colltime_dist}).

\item Halo assembly bias reflects the fact that accreting and stalled haloes populate different regions of the cosmic web.

\item Stalled haloes are more dynamically relaxed, their radial density profiles are slightly more concentrated, and a smaller fraction of their mass is in substructures with respect to the accreting haloes. 

\item Excluding their core, accreting and stalled haloes are characterized by different internal motions.
The velocity anisotropy is biased towards radial orbits in accreting haloes and towards tangential orbits
in stalled haloes. This is a consequence of the different mass-accretion modes. While matter impinges
on the accreting haloes along nearly radial orbits, the gravitational influence of the filament imparts the infalling particles a substantial tangential component for the stalled haloes. 
Due to the approximate axial symmetry of the filament, little net angular momentum is generated by this process because both senses of rotation are almost equally likely.
No obvious trend is noticeable between the collapse time and the spin parameter $\lambda_{\rm s}$ in our (small) halo sample (Table~\ref{tab_coll_a}).
The strongest signature imprinted by the filament on the halo is an enhanced tangential velocity dispersion.
Our results reveal why classifying haloes in terms of their velocity anisotropy produces a strong assembly bias \citep{Faltenbacher2010}.

\item  The supposedly universal relation between the velocity anisotropy and the slope of the radial mass-density profile found by \citet{HansenMoore2006} holds only in the innermost parts of the haloes. 
Accreting and stalled haloes show very different trends in the outer regions.

\item Accreting haloes are invariably associated with isolated density peaks in the initial conditions and their
Lagrangian patches are close to spherical. 
Stalled haloes, instead, form in between two prominent peaks. The shape of their protohaloes is very elongated along the directions perpendicular to the final filament. 
This is a consequence of the strong compressional deformation that leads to the formation of the filament.

\item The excursion-set method based on the classical spherical or ellipsoidal collapse models cannot 
reproduce the formation of the stalled haloes while it makes good predictions for the accreting ones \citepalias[see also][]{Borzyszkowski2014}.
We have presented a very simple extension of the model that accurately predicts the collapse time and the final mass of both accreting and stalled haloes. 
The key concept is that the accretion of new mass shells is inhibited whenever the asymmetric part of the tides dominates over the spherical part. 
 
\item It is common wisdom that new shells of matter continuously accrete to dark matter haloes so that they grow in mass. 
For this reason, in the calculation of the mass function \citep{Press-Schechter} or in the excursion-set model, one requires that the Lagrangian patches collapse at the same redshift at which haloes are identified, $z_{\rm id}$ \citep[although some more general formulations that release this assumption have appeared, e.g.][]{MW96,Catelanetal1998}.
Furthermore, $N$-body simulations show that the linear overdensity of protohaloes increases towards lower halo masses (this is why the halo mass function departs from the Press-Schechter form). 
The standard explanation is that lower mass haloes feel stronger tides and therefore need larger initial overdensities to collapse by $z_{\rm id}$ \citep{SMT2001}. 
Our results do not support this conjecture and provide the following alternative explanation.
Once galaxy-sized haloes are embedded in non-linear filaments of the cosmic web (a manifestation of strongly anisotropic tides), they cannot grow any longer. 
These objects must therefore assemble all their mass early on (at $z\gg z_{\rm id}$) which can only happen starting from large overdensities \citepalias[see also][]{Borzyszkowski2014}. 
\end{enumerate}

In the next two papers of this series, we will use zoom hydrodynamic simulations including star formation and feedback
to investigate the impact of the dichotomy between accreting and stalled haloes on galaxy formation and the properties of substructures.

\section*{Acknowledgements}
We thank A. Kravtsov for useful discussions and V. Springel for making {\sc pgadget3} available to us.
MB acknowledges financial support through the Transregio 33 `The Dark Universe', ERD and EG through the SFB 956 `Conditions and Impact of Star Formation' by the Deutsche Forschungsgemeinschaft.
MB thanks the Bonn-Cologne Graduate School for Physics and Astronomy for support.
The results presented were achieved employing computing resources (Cartesius) at SURF/SARA, The Netherlands as part of the PRACE-3IP project (FP7 RI-312763).

\bibliographystyle{mnras} 
 \bibliography{paper}{}  

\appendix

\label{lastpage}

\end{document}